\documentclass[fleqn,usenatbib]{mnras}

\usepackage{newtxtext}
\usepackage{mathptmx}
\usepackage{comment}

\usepackage[T1]{fontenc}

\DeclareRobustCommand{\VAN}[3]{#2}
\let\VANthebibliography\thebibliography
\def\thebibliography{\DeclareRobustCommand{\VAN}[3]{##3}\VANthebibliography}

\usepackage{graphicx}	
\usepackage{amssymb}	
\usepackage{amsmath}	

\def\aj{AJ}             	
\def\araa{ARA\&A}       	
\def\apj{ApJ}           	
\def\apjl{ApJ}          	
\def\apjs{ApJS}         	
\def\aap{A\&A}          	
\def\mnras{MNRAS}       	

\title[Cosmic-structure classification for galaxies]{Classification of cosmic structures for galaxies with deep learning: connecting cosmological simulations with observations}

\author[S. Inoue et al.]{
{Shigeki Inoue$^{1}$, Xiaotian Si$^{1}$, Takashi Okamoto$^{1}$ \& Moka Nishigaki$^{2,3}$}
\\
$^{1}$Faculty of Science, Hokkaido University, Sapporo, Hokkaido 060-0810, Japan\\
$^{2}$Department of Astronomical Science, SOKENDAI (The Graduate University for Advanced Studies), 2-21-1 Osawa, Mitaka, Tokyo, Japan\\
$^{3}$National Astronomical Observatory of Japan, National Institute of Natural Sciences, 2-21-1 Osawa, Mitaka, Tokyo, Japan
}

\date{Accepted XXX. Received YYY; in original form ZZZ}

\pubyear{2022}

\begin{document}
\label{firstpage}
\pagerange{\pageref{firstpage}--\pageref{lastpage}}
\maketitle

\begin{abstract}
We explore the capability of deep learning to classify cosmic structures. In cosmological simulations, cosmic volumes are segmented into voids, sheets, filaments and knots, according to distribution and kinematics of dark matter (DM), and galaxies are also classified according to the segmentation. However, observational studies cannot adopt this classification method using DM. In this study, we demonstrate that deep learning can bridge the gap between the simulations and observations. Our models are based on three-dimensional convolutional neural networks and trained with data of distribution of galaxies in a simulation to deduce the structure classes from the galaxies rather than DM. Our model can predict the class labels as accurate as a previous study using DM distribution for the training and prediction. This means that galaxy distribution can be a substitution for DM for the cosmic-structure classification, and our models using galaxies can be directly applied to wide-field survey observations. When observational restrictions are ignored, our model can classify simulated galaxies into the four classes with an accuracy (macro-averaged $F_{\rm 1}$-score) of 64 per cent. If restrictions such as limiting magnitude are considered, our model can classify SDSS galaxies at $\sim100~{\rm Mpc}$ with an accuracy of 60 per cent. In the binary classification distinguishing void galaxies from the others, our model can achieve an accuracy of 88 per cent. 


\end{abstract}




\section{Introduction}
\label{Intro}
The formation and evolution of galaxies can depend on their environments in the Universe. Generally, galaxies in dense regions such as galaxy clusters are thought to form earlier than those in underdense regions such as voids. In clusters, various interactions between galaxies can affect their properties, such as tidal effects, mergers, gas stripping and metal pollution by outflows from other galaxies \citep[e.g.][and references therein]{d:80,bm:09}. On the other hand, field galaxies outside the clusters can be expected to accrete inter-galactic medium and may sustain their star formation activity until relatively low redshifts \citep[e.g.][]{ans:19}. \citet{dtk:22} have found that spatial gradients of metallicities are significantly weaker in filaments than those in clusters. To understand the formation and evolution of galaxies, it is therefore indispensable to accurately know which cosmic structures galaxies belong to.

It is, however, not straightforwards to classify galaxies into cosmic structures. Large-scale structures in the Universe are formed by the gravity of dark matter (DM). Therefore, theoretical studies often utilise cosmological simulations and categorise spatial regions according to the topology of local DM density fields (see also Section \ref{sim_csc}) into four classes: void, sheet, filament and knot \citep[e.g.][]{hpc:07,ajv:07,fhg:09,hmy:12}.\footnote{The sheet and knot are also referred to as `wall' and `node'.} The knots are converging regions of DM density, thought to correspond to galaxy clusters. The filaments are one-dimensional segments connecting the knots and embedded in two-dimensional structures of sheets. The voids are nearly static or expanding regions with low densities. However, since these theoretical methods rely on the distribution of DM particles in simulations, they cannot be directly applied to actual observations of galaxies.

Observational methods of cosmic-structure classification generally use the spatial distribution of galaxies, instead of DM, obtained by surveys for vast fields such as the Sloan Digital Sky Survey (SDSS). The observational classification based on galaxies is performed with various methods depending on purposes of the studies. For example, the method of {\sc DisPerSE} \citep[Discrete Persistent Structures Extractor, ][]{spc:08,s:11} extracts filaments by connecting local density peaks and saddle points of the galaxy distribution. \citet{tsm:14,tks:14} also identify filament regions as concatenated cylinders with a constant width. Voids are often defined as regions centred on minima of galaxy density with outer boundaries determined by a watershed algorithm or maximum spheres devoid of galaxy \citep[e.g.][]{hv:02,kpa:11,lw:12,slw:12,hpg:20}. Most of the observational studies generally classify galaxies into three at most: void, filament and knot (cluster).

In terms of methodology, the theoretical and observational classification of cosmic structures thus relies on the different components: DM and observable galaxies. Because the formation of cosmic structures is driven by self-gravity of DM as we mention above, the theoretical classification based on DM is considered to be plausible; however DM is not observable. Although the observational classification is based on galaxy distribution, it is not guaranteed that galaxies can be taken as accurate tracers of DM; note that the formation of galaxies can depend on environments. Observations cannot detect faint galaxies below their limiting magnitudes. In addition, star formation does not occur in low-mass haloes below a threshold of $M_{\rm DM}\sim10^{9.5}~{\rm M_\odot}$ at redshift $z\lesssim1$ \citep[e.g.][]{e:92,okamoto08a,bf:20}. Therefore, the observational classification based on galaxy distribution may be unable to capture details of the cosmic structures in low-density environments.

To address the problem mentioned above, we propose a novel method that is based on deep learning and applicable to observations. \citet{a:19} has demonstrated the ability of artificial neural networks to classify the cosmic structures using DM density distribution in a cosmological $N$-body simulation (see also Section \ref{gridbase}). In this study, we utilise a cosmological simulation including both DM and baryons, in which we can simultaneously access the cosmic-structure classification obtained by the DM-based analysis and spatial distribution of observable galaxies hosting stars. Our model learns the relationship between the class labels and observable quantities such as galaxy number count and predicts the cosmic structures using the galaxy distribution instead of DM. The model trained with the simulated galaxies can be applied to real observations such as SDSS.

This paper is structured as follows. In Section \ref{sim_csc}, we describe the cosmological simulation we utilise and the analysis of DM for the cosmic-structure classification. In Section \ref{ml_model}, we explain a layout of our neural networks and creation of our learning data to train our model. First, in Section \ref{gridbase}, we evaluate the intrinsic performance of our model and compare it with a previous study. In Section \ref{galwithgridbase}, we argue the difference of sampling schemes for training data and its influence on our deep-learning models. Next, in Section \ref{galbase}, we move on to the application to our mock observations and estimate the expected accuracy of our model for observational data. As the goal of this paper, in Section \ref{appSDSS}, we classify observed galaxies in the SDSS data with our models. We discuss our results in Section \ref{discussion} and draw conclusions from this study in Section \ref{summary}. 

\section{The cosmological simulation and cosmic structure classification}
\label{sim_csc}
We use data sets of the IllustrisTNG project. The details of the cosmological simulations are presented on the website\footnote{\url{https://www.tng-project.org}} and in related papers including \citet{TNG}, \citet{wsh:17} and \citet{psn:18}. The simulations are performed with the $N$-body/moving-mesh hydrodynamics code {\sc Arepo} \citep{arepo,wvr:20} in which various sub-resolution physics such as gas cooling, star and black hole formation, and their feedback effects are implemented. This study focuses on a snapshot of TNG100-1 at redshift $z=0$; we also analyse the higher-resolution run, TNG50-1, in Appendix \ref{TNG50}. The simulation box of TNG100-1 has a comoving side length of $111~{\rm Mpc}$, and the mass-resolutions of DM and stellar particles are $7.5$ and $\simeq1.4\times10^6~{\rm M_\odot}$, respectively. The cosmological parameters of the total matter, dark energy and baryonic densities are $\Omega_{\rm m}=0.3089$, $\Omega_{\rm \Lambda}=0.6911$ and $\Omega_{\rm b}=0.0486$, respectively. The Hubble constant $H_0=67.74~{\rm km~s^{-1}~Mpc^{-1}}$ is adopted. Gravitationally bound structures are identified with the friends-of-friends and {\sc SUBFIND} grouping algorithms \citep[e.g.][]{swt:01}. In this study, the total masses of stars and DM are computed for each {\sc SUBFIND} group which is considered to be a single (sub)halo.\footnote{We do not distinguish between main and subhaloes, refer to both as haloes in this study.} Stellar magnitudes in mock SDSS bands are computed with a photometric model of \citet{bc:03}.

We use the method proposed by \citet{hmy:12} for the cosmic-structure classification. First, we assign mass of the DM particles in the whole simulation to $256^3$ voxels with the cloud-in-cell algorithm, which means that our classification has a spatial resolution of $\Delta r=432~{\rm kpc}$. We do not use stars or gas for this analysis. Then, we obtain the mean velocities of DM in the voxels and adopt a Gaussian smoothing with the kernel size of $\Delta r$ to the velocity fields in order to wash out voxel-scale noise. Next, we apply fast Fourier transform and compute tensors of velocity gradients as 
\begin{equation}
    \Sigma_{ij} = -\frac{1}{2H_0}\left(\frac{\partial v_i}{\partial r_j} + \frac{\partial v_j}{\partial r_i}\right),
    \label{VelGraTendor}
\end{equation}
where the subscripts of $i$ and $j$ represent $x$, $y$ and $z$ in the Cartesian coordinates. Finally, we compute the three eigenvalues of the tensor $\Sigma_{ij}$ and count the number of the eigenvalues larger than a threshold $\lambda_{\rm th}=0.44$.\footnote{This threshold is the value recommended in \citet{hmy:12} according to their visual inspection.} When all, two, one and none of the eigenvalues is larger than $\lambda_{\rm th}$, the voxel is categorised into knot, filament, sheet and void, respectively. To remedy over-resolutions, we apply the multi-scale approach for the high-density regions above certain thresholds \citep[see section 5 of ][]{hmy:12}; in short we recompute the tensors by adopting a broader Gaussian kernel with $4\times\Delta r$, and replace the categorisation if the results change.

\begin{figure}
  \includegraphics[bb=0 0 1476 2851, width=\hsize]{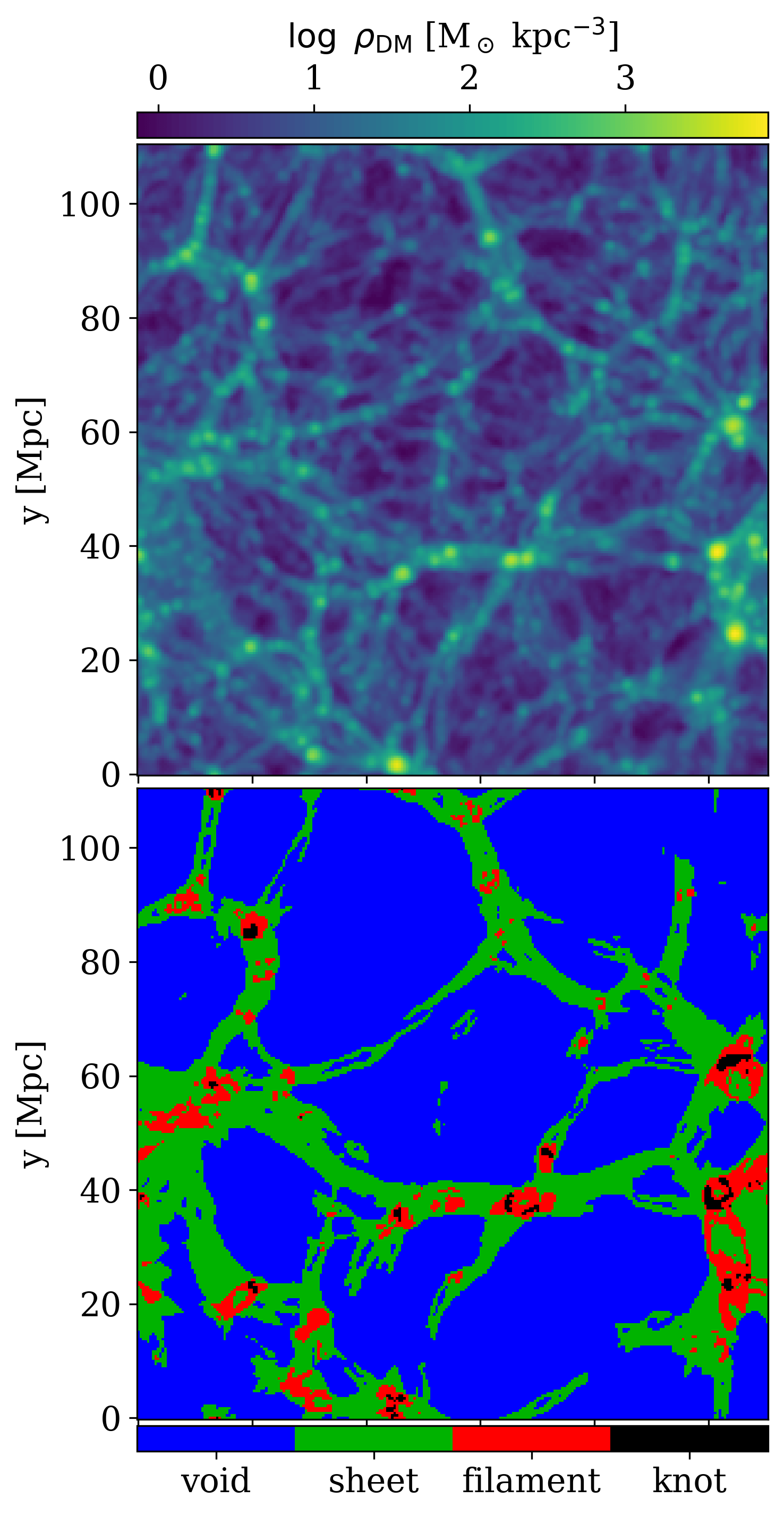}
\caption{\textit{Top}: DM densities in a slice as thin as $\Delta r=432~{\rm kpc}$ in TNG100-1 at redshift $z=0$, where the density fields are smoothed with the voxel-size kernel with $\Delta r$. \textit{Bottom}: the grid-based classification of the cosmic structures on the slice.}
\label{DMmap}
\end{figure}
Fig. \ref{DMmap} shows our result of the cosmic-structure classification described above. In comparing the DM density map (top), regions classified as sheets distribute along structures resembling webs in the bottom panel. This is because the figure shows cross-sections of two-dimensional structures of the sheets. In the bottom panel, most of the voxels classified as filaments are found in intersections of the sheets, and knots are surrounded by the filament voxels. In the bottom panel, some weak structures of sheets are disconnected and/or `hollow' in their cores, and these regions are classified as voids. This may be indicative of the inaccuracy of the above analysis. The sensitivity to detect such weak structures would depend on the values of $\lambda_{\rm th}$ and $\Delta r$. In this study, we take a stance that the above analysis based on \citet{hmy:12} gives the definition of the cosmic structures although there could be mis-labelling due to such inaccuracy of the analysis. We argue the influence by the details of the DM analysis in Appendix \ref{mislabel}. This analysis classifies the equally spaced grid points representing $256^3$ voxels. In this simulation, we find that $68.1$, $26.7$, $4.83$ and $0.344$ per cent of the voxels are classified as voids, sheets, filaments and knots, respectively. In this paper, we hereafter refer to this classification as `grid-based classification'.

\begin{figure}
  \includegraphics[bb=0 0 2539 2460, width=\hsize]{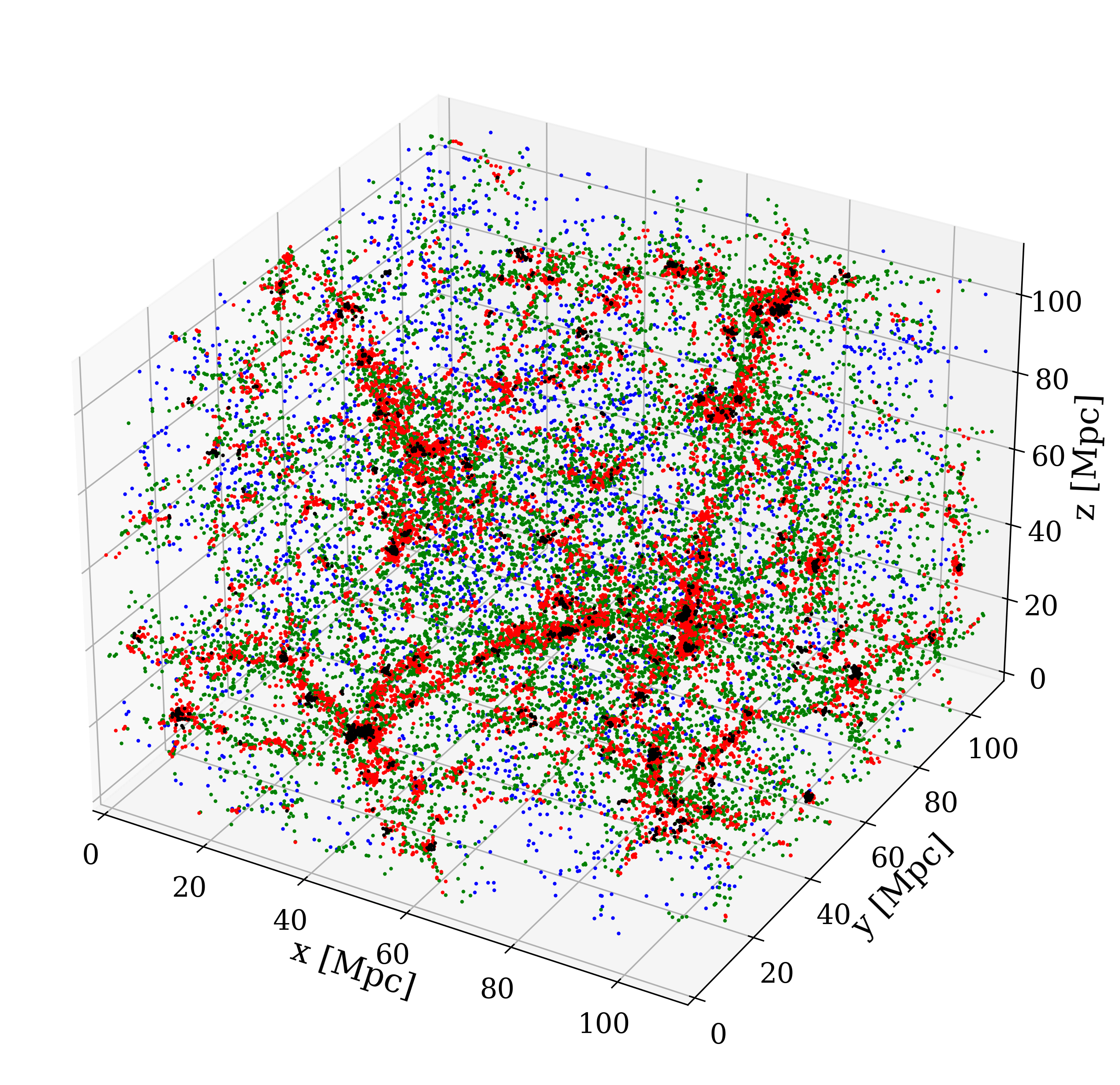}
  
  \begin{minipage}[b]{0.5\linewidth}
    \centering
  \includegraphics[bb=0 0 2460 2460, width=\linewidth]{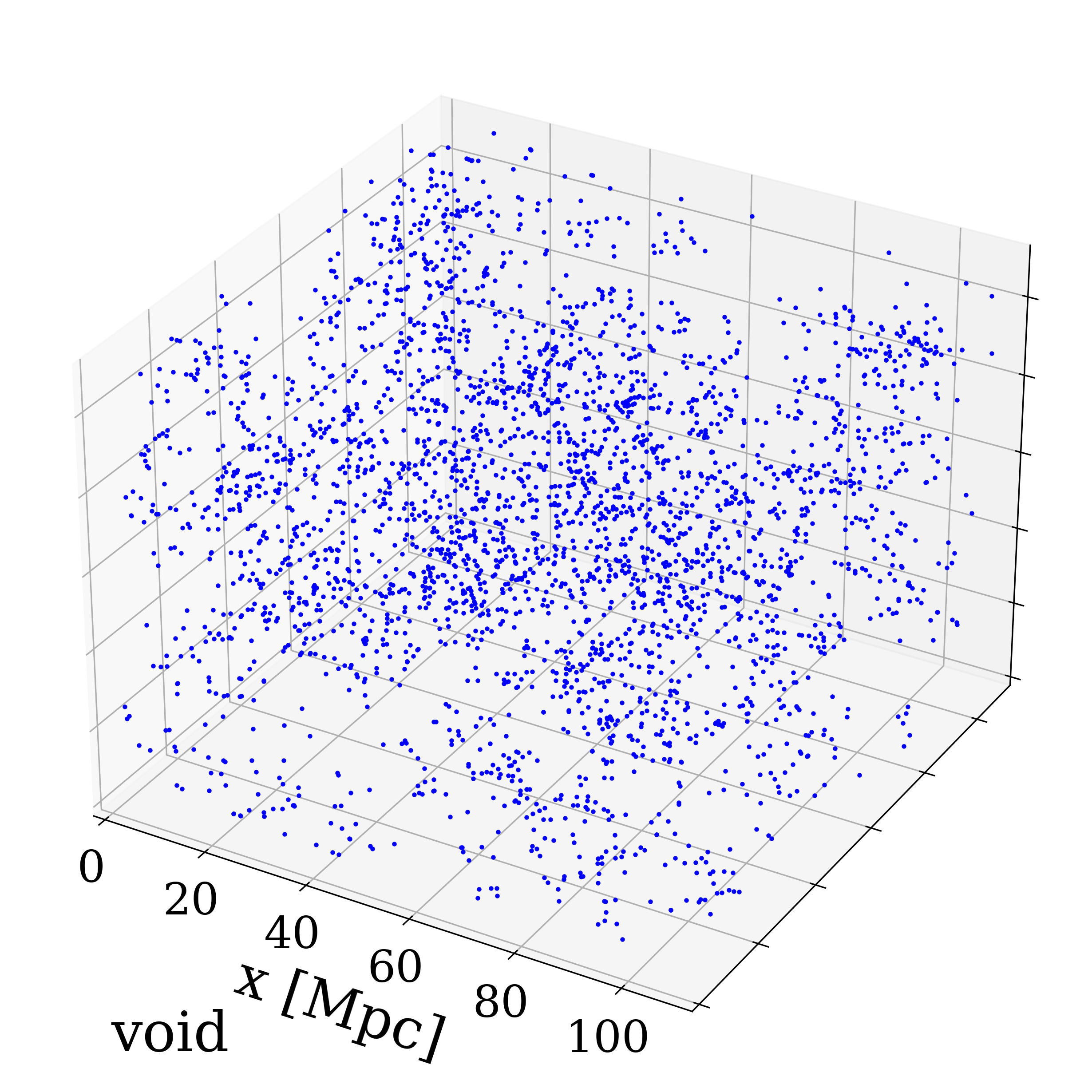}
  \end{minipage}
  \begin{minipage}[b]{0.5\linewidth}
    \centering
  \includegraphics[bb=0 0 2460 2460, width=\linewidth]{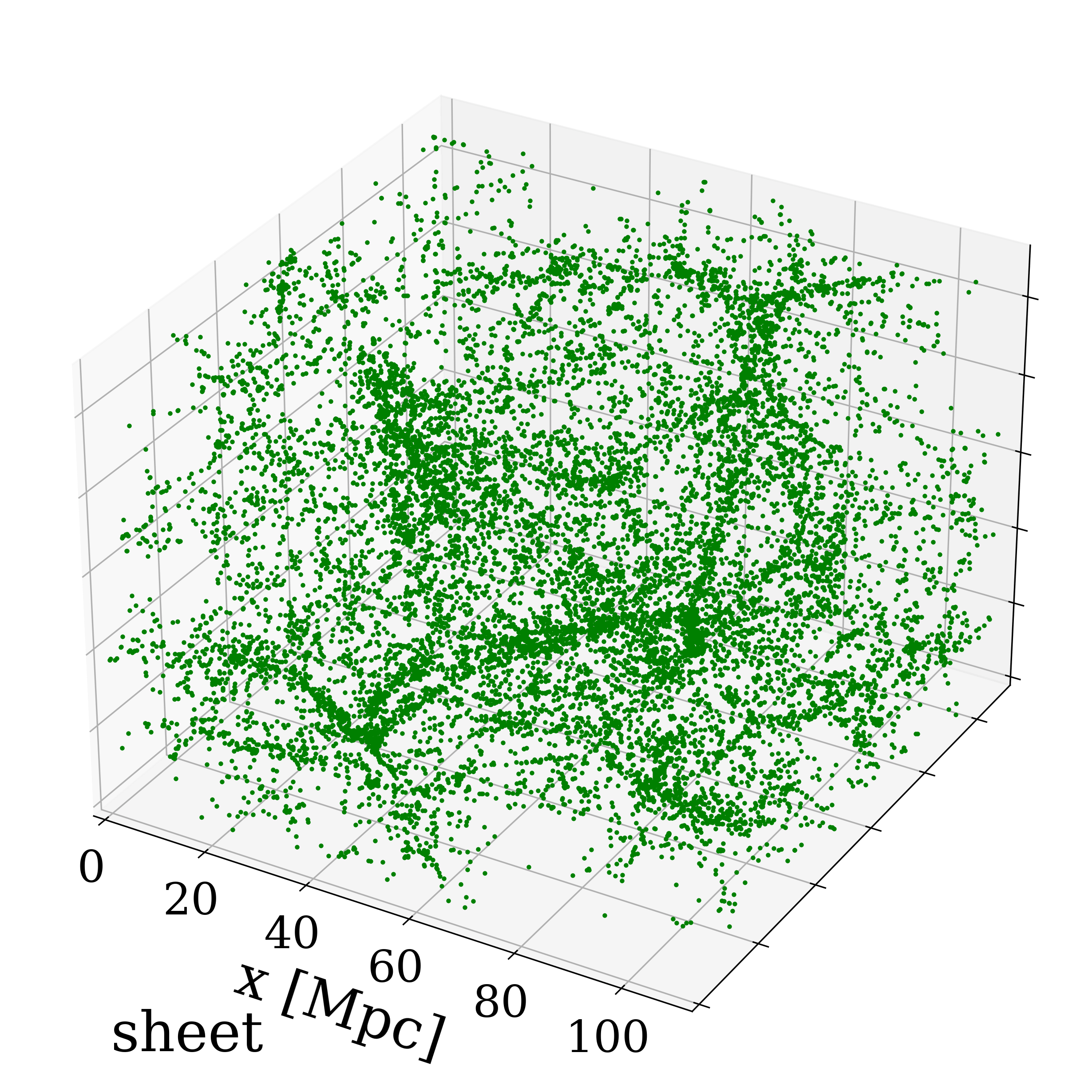}
  \end{minipage}
  \begin{minipage}[b]{0.5\linewidth}
    \centering
  \includegraphics[bb=0 0 2460 2460, width=\linewidth]{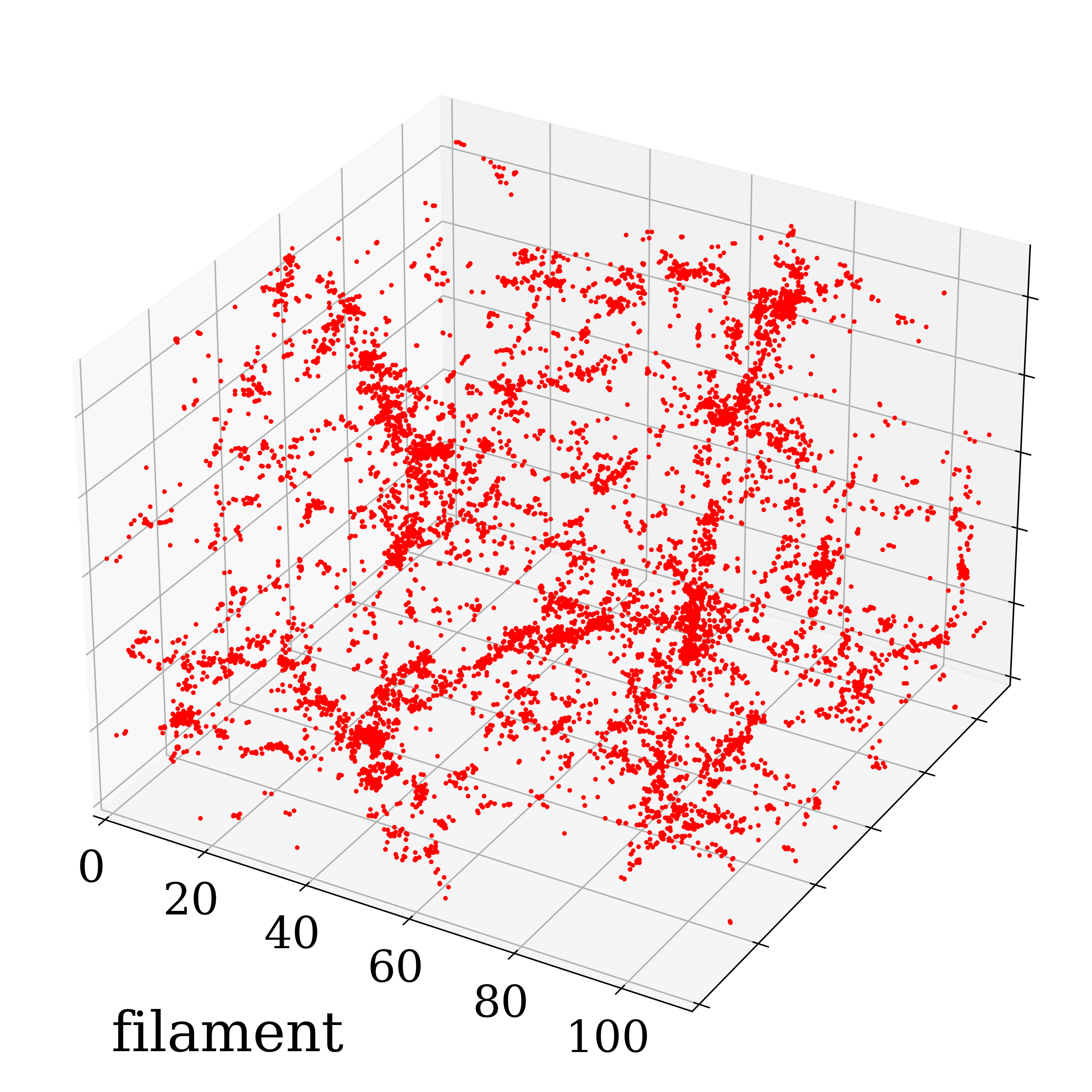}
  \end{minipage}
  \begin{minipage}[b]{0.5\linewidth}
    \centering
  \includegraphics[bb=0 0 2460 2460, width=\linewidth]{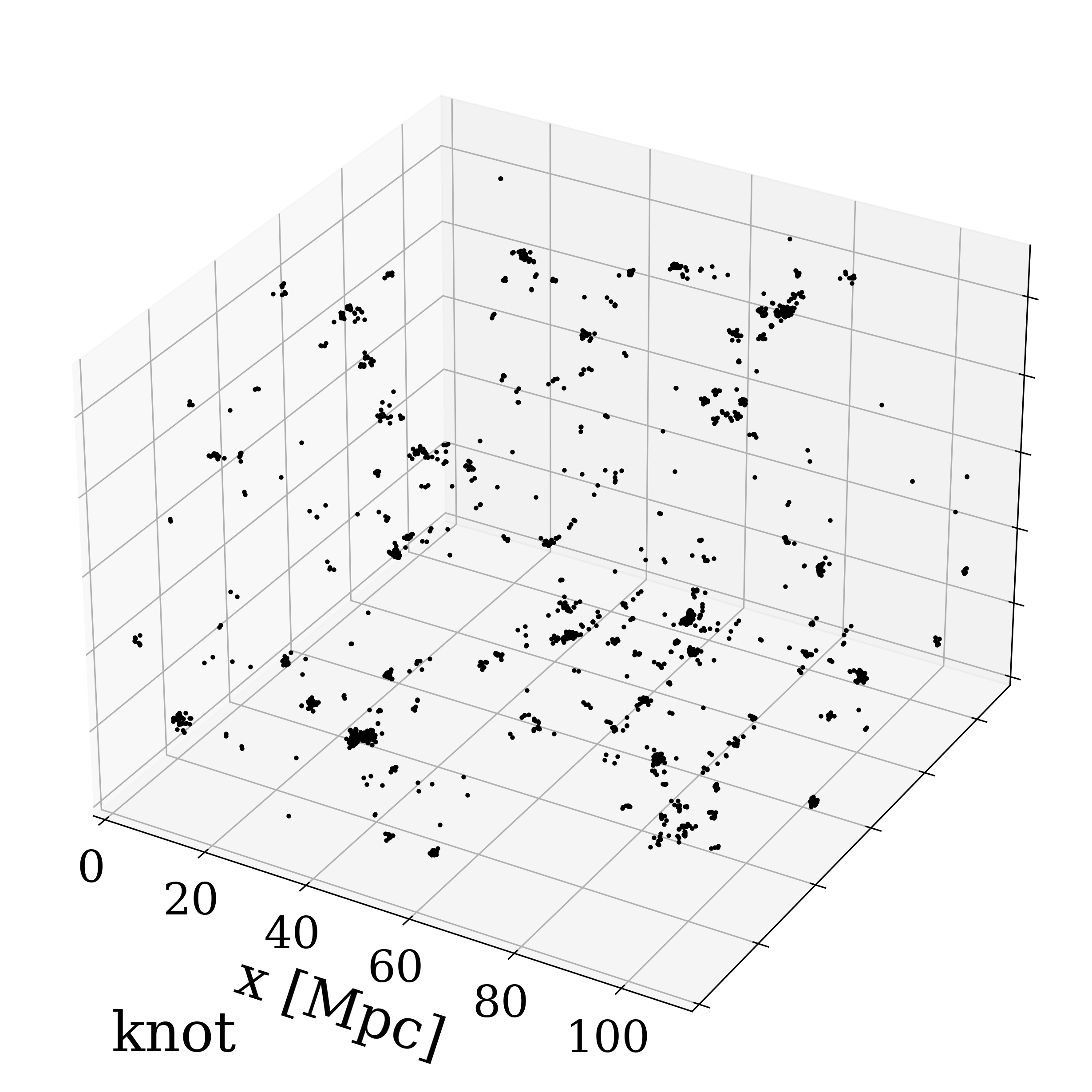}
  \end{minipage}
\caption{Three-dimensional distribution of the galaxies labelled as voids (blue), sheets (green), filaments (red) and knots (black) in TNG100-1 at redshift $z=0$: the galaxy-based classification. We here plot galaxies with stellar masses of $M_{\rm star}>10^{8.5}~{\rm M_\odot}$ for visibility. The filament galaxies appear to form web structures connecting the knots, just like neurons. The bottom set of panels plots the same galaxies but separately for each label.}
\label{CosWeb}
\end{figure}

\begin{figure}
  \includegraphics[bb=0 0 1866 1169, width=\hsize]{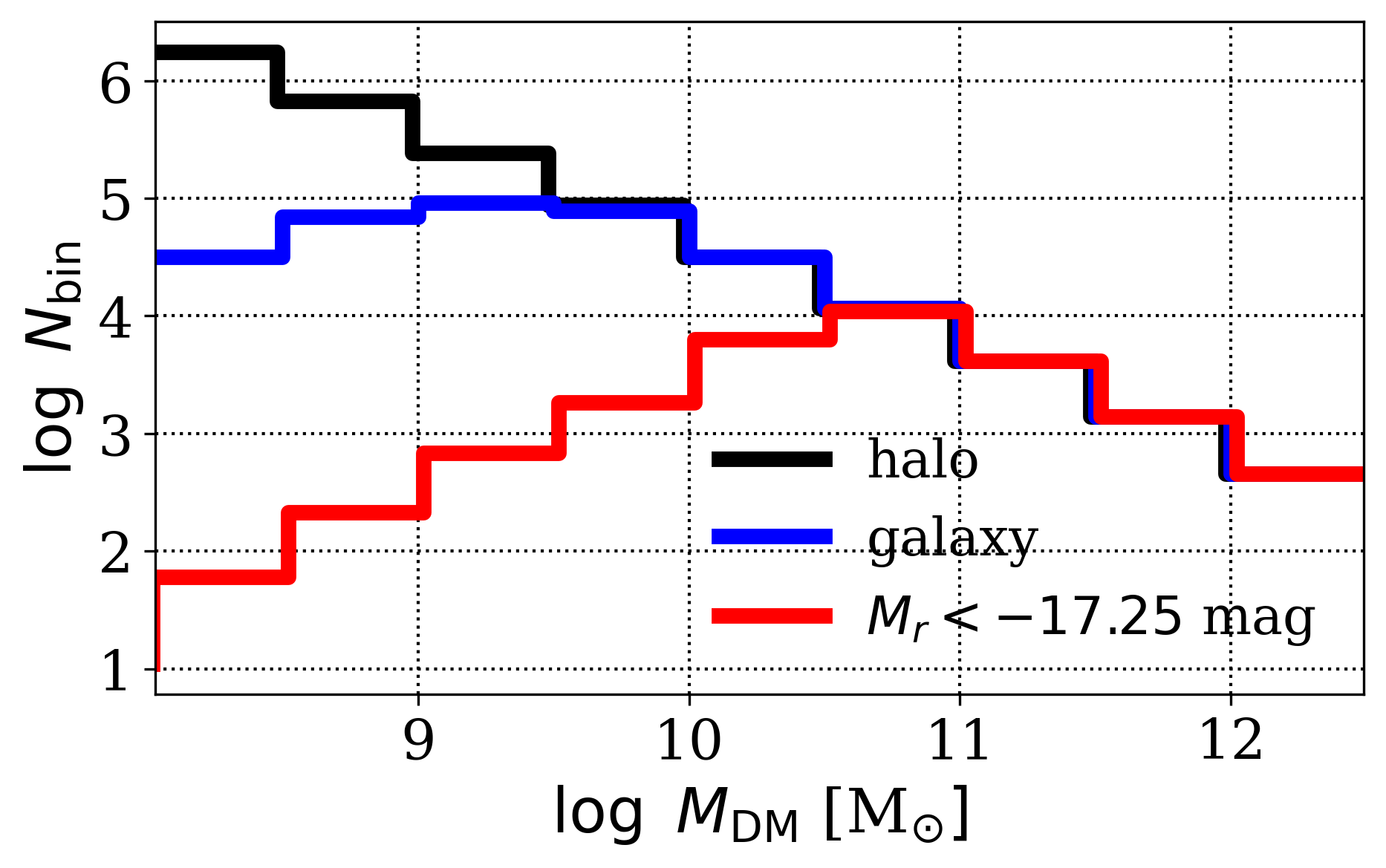}
\caption{Histograms of the number of {\sc SUBFIND} groups as functions of DM mass. The black line indicates all (sub)haloes in TNG100-1 at redshift $z=0$, and the blue line does those hosting stars. The red line is for galaxies brighter than an $r$-band absolute magnitude of $M_r=-17.25~{\rm mag}$ which corresponds to the apparent magnitude limit of SDSS, $m_r=17.75~{\rm mag}$, assuming the galaxies to be at a distance of $100~{\rm Mpc}$ (see Section \ref{galbase} and Appendix \ref{SDSSlimit}). The halo occupation fractions are close to unity above $M_{\rm star}\sim10^{9.5}~{\rm M_\odot}$ and $M_{\rm star}\sim10^{10.5}~{\rm M_\odot}$ for galaxies having stars (blue) and those brighter than the limiting magnitude (red).}
\label{HOF}
\end{figure}
This study aims to classify galaxies rather than the spatial voxels. We label each galaxy with the classification of the voxel in which the galaxy resides. Fig. \ref{CosWeb} shows the three-dimensional positions of the galaxies coloured with their labels. The galaxies labelled as filaments (red) appear to distribute like strings connecting those labelled as knots (black). In the three-dimensional distribution, our classification thus appears to represent well the cosmic webs in the simulation. If we define `galaxies' as SUBFIND groups having at least one stellar particle in the simulation, we find that $45518$, $152534$, $105291$ and $22390$ galaxies are labelled as voids, sheets, filaments and knots, respectively. Hereafter, we refer to this classification as `galaxy-based classification'. Fig. \ref{HOF} (blue) shows the mass distribution of galaxies with stars, and most of the {\sc SUBFIND} groups above $\sim10^{9.5}~{\rm M_\odot}$ in DM mass host stars. Note that the halo masses of the galaxies that marginally form stars can depend on resolution and sub-grid models in the simulation (see Appendix \ref{TNG50}). If we consider galaxies brighter than an absolute magnitude of $M_r=-17.25~{\rm mag}$ in $r$-band, these galaxies occupy most of the haloes above $\sim10^{10.5}~{\rm M_\odot}$ (see Section \ref{galbase}).

\section{Constructing machine learning models}
\label{ml_model}
\subsection{Creating cubic data}
\label{CreateData}
\begin{figure*}
  \includegraphics[bb=0 0 2743 1133, width=\hsize]{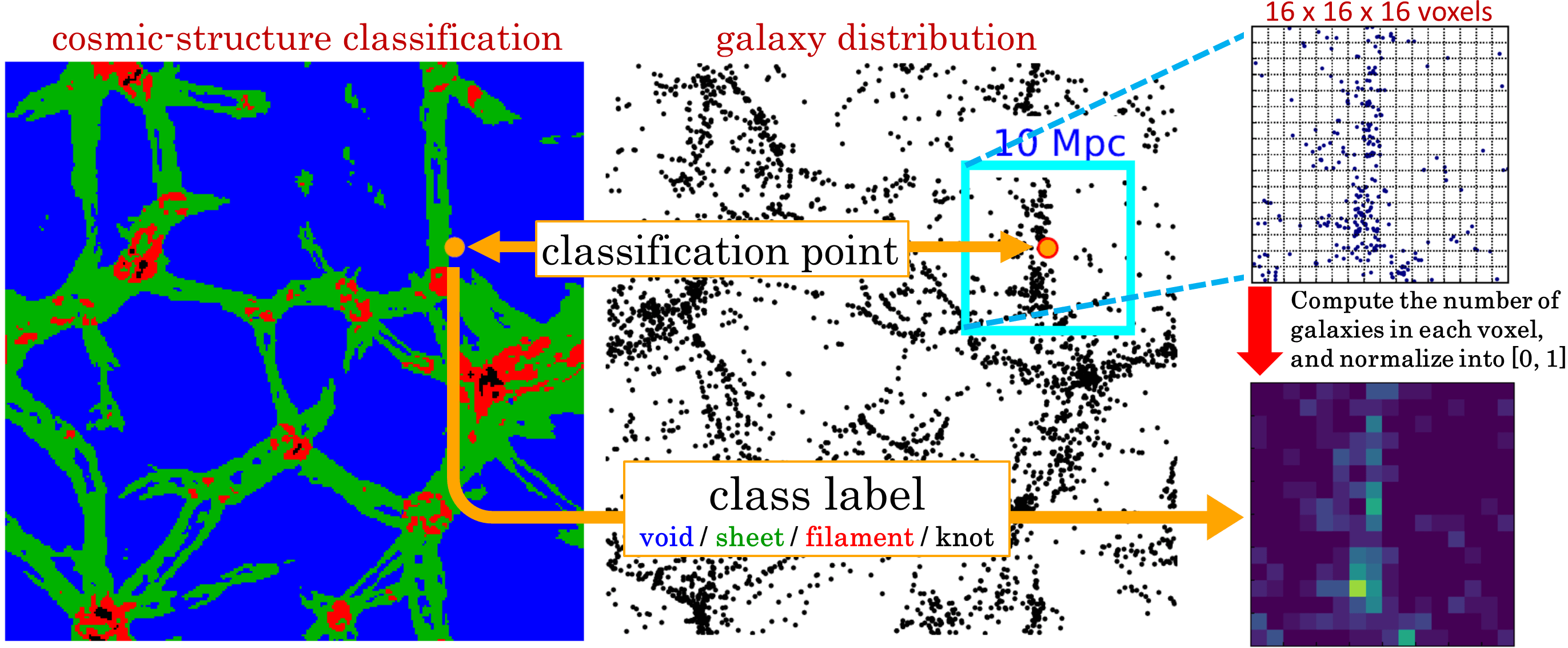}
\caption{Schematic illustration of creating our learning data and assigning them class labels (see Section \ref{CreateData}). Note that our actual data are in three-dimensional cubic volumes although this figure illustrates the sampled region of data in the two-dimensional manner for simplicity. The left panel shows a map of cosmic-structure classification like the bottom one of Fig. \ref{DMmap}, and the central panel indicates the distribution of galaxies in the same region. A classification point (the orange filled circles) is randomly selected from the grid points or the galaxies in the simulation, and a cubic region of $10^3~{\rm Mpc^3}$ centred on the classification point is randomly oriented (the cyan square in the central panel), within which galaxies are sampled. In the sampled region, the numbers of galaxies are computed for $16^3$ voxels without any smoothing (top right panel), and the cubic data are normalised among all the created data (bottom right). The data cube takes over the label of its classification point.}
\label{schematic}
\end{figure*}
To feed our model in Section \ref{CNN}, we create data of three-dimensional cubic regions of galaxy distribution tagged with the labels of the cosmic structures. We set the side length of all cubic data to be $10~{\rm Mpc}$, and a cube is binned with $16^3$ voxels whose size is $625~{\rm kpc}$. Each cube is centred on its `classification point' assigned with the labels obtained in Section \ref{CosWeb}: knot, sheet, filament or void. A position of the classification point is selected at random from the grid points in the case of the grid-based classification, whereas it is sampled from the galaxies in the galaxy-based classification. Next, we rotate all galaxies three-dimensionally around the normal, transverse and longitudinal axes at random while fixing the classification point. Then, galaxies in the cubic region are placed in the voxels according to their positions. In the galaxy-based classification, the galaxy selected as the classification point is also included in the data. Here we do not apply the could-in-cell algorithm or any other smoothing schemes since we find that applying such a smoothing does not improve the performance of our models. Finally, we tag the cubic data with the class label at the classification point. Fig. \ref{schematic} illustrates the above procedures schematically.

The side length of the cubic region, $10~{\rm Mpc}$, is set arbitrarily. We find, however, that our results hardly change in the range from $\sim10$ to $20~{\rm Mpc}$ and become less accurate outside this range. The number of voxels in the cube, $16^3$, is also an arbitrary choice; however we confirm that this little affects the accuracy of our models between $8^3$ and $32^3$.

Since the grid- and galaxy-based classification are different in the schemes to sample the classification points, the data produced by the two schemes are not qualitatively homogeneous. Because the distribution of galaxies is biased towards high-density environments, a single cubic region in the galaxy-based classification generally includes more galaxies than that in the grid-based classification even if their labels are the same. The three-dimensional extension of galaxy distribution could also be different between them. We argue the influence on our 3D-CNN models by these different sampling schemes in Section \ref{galwithgridbase}.

\subsection{data augmentation and preprocessing}
\label{DataAug}
By repeating the above procedures, we create 10000 cubic data for each label. The classification points are randomly sampled with replacement. Although the same classification points can be selected again, we randomly change the orientation of the cube every time. In this study, the cubic data consist of a single channel: the number of galaxies, $N_{\rm voxel}$, in each voxel. We scale $N_{\rm voxel}$ linearly in the data. Even when we scale the data logarithmically, the performance of our models hardly changes.\footnote{We give the voxel $\log(N_{\rm voxel}+1)$ in the logarithmic case.} We discuss the influence by including other physical quantities as additional channels in Section \ref{improve}. For preprocessing the data, we apply the Min-Max normalisation. The maximum value is computed among all voxels in all cubes, and then all voxel values are normalised into $[0, 1]$. In the 10000 data cubes for each label, 6400, 1600 and 2000 are used as training, validation and test data. Even if we generate four times more data sets, the performance of our models is not improved.

\subsection{Neural networks: simple 3D-CNN as classifier}
\label{CNN}
We expect that the differences in three-dimensional spatial distribution of ambient galaxies would have key information in deducing the cosmic structures. We, therefore, consider three-dimensional neural networks (3D-CNN) to be a feasible classifier of the cosmic structures. Such a classifier would be able to detect and distinguish their diffuse (voids), planar (sheets), filamentary (filaments) and concentrated (knots) distribution of galaxies.

We implement 3D-CNN using the {\sc Keras} library on {\sc TensorFlow}. Our network architecture is shown in Table \ref{layers}. We adopt the Adamax optimiser \citep{Adam} with a learning rate of $0.002$ without decaying and the parameters of $\beta_1=0.9$, $\beta_2=0.999$ and $\varepsilon=10^{-7}$. The categorical cross-entropy loss function is used. We set a mini-batch size to $512$ and the number of learning epochs $N_{\rm epoch}$ to $30$. We use the model trained at the last epoch $N_{\rm epoch}=30$ to evaluate the resultant accuracy of the model using the test data; the performance of our models does not significantly depend on the epoch when we stop the training as long as $N_{\rm epoch}\gtrsim10$ (see Section \ref{results}). 
\begin{table}
  \begin{center}
    \leavevmode
\begin{tabular}{l c c c c} 
\hline
\hline
layer type    & kernel size       & $N_{\rm filter}$ / $N_{\rm out}$  & remarks\\
\hline 
Convolution 1 & $3\times3\times3$ & $N_{\rm filter}=32$ & ReLU / He uniform\\
Maxpool 1     & $2\times2\times2$ & -  & - \\
Dropout 1     &       -           & -  & $R_{\rm drop}=0.25$\\[0.05in]

Convolution 2 & $3\times3\times3$ & $N_{\rm filter}=64$ & ReLU / He uniform\\
Maxpool 2     & $2\times2\times2$ & -  & - \\
Dropout 2     &       -           & -  & $R_{\rm drop}=0.25$\\[0.05in]
Flatten       &       -           & -  & - \\[0.05in]
Fully connected 1 & - &$N_{\rm out}=512$& ReLU / He uniform\\
Fully connected 2 & - &$N_{\rm out}=512$& ReLU / He uniform\\
Output            & - &$N_{\rm out}=N_{\rm class}$& softmax\\
\hline 
\hline
  \end{tabular} 
\caption{Layers in our 3D-CNN architecture. The third column indicate the numbers of filters $N_{\rm filter}$ and output units $N_{\rm out}$ of convolution and fully connected layers. The fourth column describes parameters and types of activation function and initialiser, where $R_{\rm drop}$ is dropout rate of the max-pooling layers, and ReLU stands for Rectified Linear Unit.}
  \label{layers}
  \end{center}
\end{table}

We find that such a simple 3D-CNN architecture with only two convolutional layers works well enough without significant overfitting in this study (see below), and building deeper networks can result in a tendency of instability on loss values as well as overfitting. Although we have tested various architectures, hyper-parameters, other machine/deep-learning libraries and ways of preprocessing, such alteration for our model only changes the total accuracy by $\lesssim0.03$ (see Section \ref{improve}). Hence, we do not change the architecture or hyper-parameters throughout this paper unless otherwise stated. 

\section{Results}
\label{results}
\subsection{The grid-based classification}
\label{gridbase}
\begin{figure}
  \includegraphics[bb=0 0 769 493, width=\hsize]{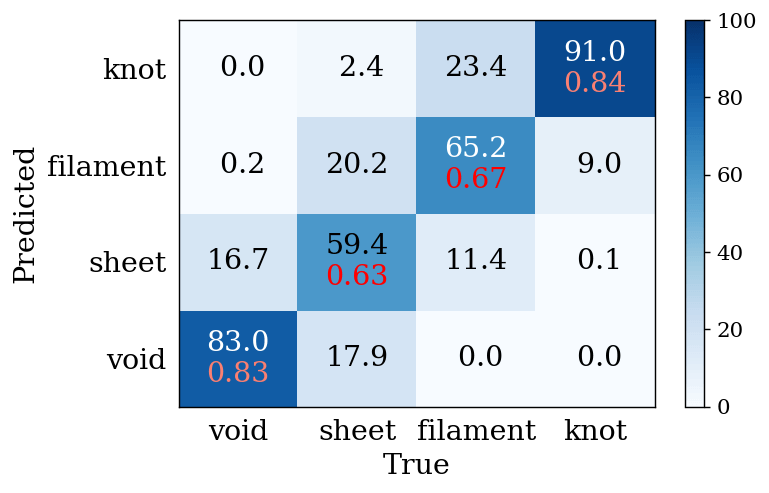}
\caption{The normalised confusion matrix of the grid-based classification into the four categories. The value written with black or white in each cell indicates a percentage of the prediction for the test data. For example, for the two bottom cells in the second column from the left, $59.4$ and $17.9$ per cent of the cubic data labelled as `sheet' are classified correctly as `sheet' and erroneously as `void', respectively. The sum in each column is $100$ per cent. The values written with red  in the diagonal cells indicate $F_{\rm 1}$-scores of the classes. The macro-average of the $F_{\rm 1}$-scores is $0.74$ for the test data.}
\label{CM_grid_four}
\end{figure}
We first examine the intrinsic performance of our 3D-CNN models. For this purpose, we present our results for the grid-based classification and compare them with a previous study. We here randomly select the classification points of the cubic data from the $256^3$ grid points, and number densities in the cubes are computed from the halos containing stellar particles. Fig. \ref{CM_grid_four} tabulates the normalised confusion matrix and $F_{\rm 1}$-scores in the case of classification into the four categories. The $F_{\rm 1}$-score is equivalent to Dice coefficient in the case of binary classification, defined as 
\begin{equation}
    F_{\rm 1}\equiv\frac{N_{\rm tp}}{N_{\rm tp}+\frac{1}{2}\left(N_{\rm fp}+N_{\rm fn}\right)},
\end{equation}
where $N_{\rm tp}$, $N_{\rm fp}$ and $N_{\rm fn}$ are the numbers of true positive, false positive and false negative classification. In this study, we evaluate the performance of our models with the $F_{\rm 1}$-scores. As shown in Fig. \ref{CM_grid_four}, the model can achieve relatively high accuracy for voids and knots with the $F_{\rm 1}=0.83$ and $0.84$. On the other hand, classifying sheets and filaments are less accurate with the lower $F_{\rm 1}$-scores of $0.63$ and $0.67$. In the two classes, $20.2$ and $17.9$ per cent of the sheet voxels are erroneously classified as filaments and voids, and $23.4$ and $11.4$ per cent of the filament voxels are mistaken for knots and sheets. This difficulty in distinguishing sheets and filaments is probably because of their intermediate densities and complex spatial extension (see also Section \ref{locden}). However, the macro-average of the $F_{\rm 1}$-scores is $0.74$,\footnote{In all computations in this paper, the macro-averages of $F_{\rm 1}$-scores are little different from the micro-averages. The macro-average is defined as the mean of $F_{\rm 1}$ computed in each class, whereas the micro-average is defined as $F_{\rm 1}$ computed with the means of $N_{\rm tp}$, $N_{\rm fp}$ and $N_{\rm fn}$ among the classes.} which means that our model is not useless although we need to be aware of the accuracy according to actual applications of this method.

\begin{figure}
  \includegraphics[bb=0 0 806 500, width=\hsize]{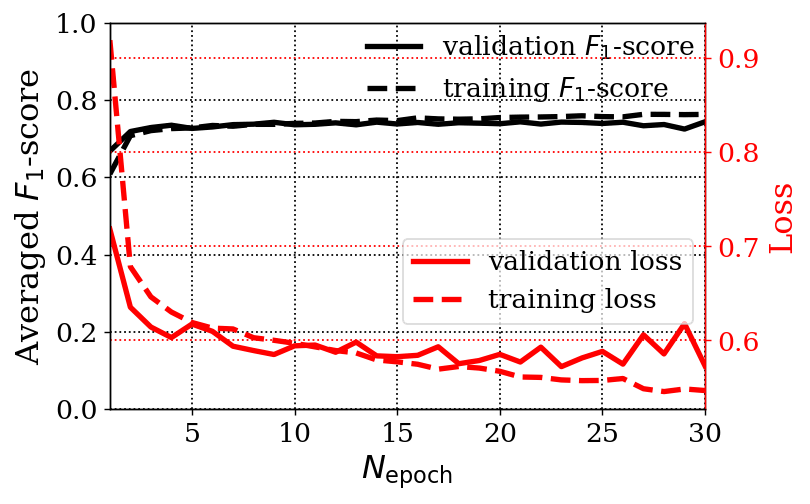}
\caption{The macro-averaged $F_{\rm 1}$-scores (the black lines with the left ordinate) and loss values (the red lines with the right ordinate) as functions of learning epoch $N_{\rm epoch}$ in the case of Fig. \ref{CM_grid_four}: the grid-based classification into the four categories. The solid and dashed lines indicate the values for the validation and training data.}
\label{LA_grid_four}
\end{figure}
For the above quaternary classification, Fig. \ref{LA_grid_four} indicates the macro-averages of $F_{\rm 1}$-scores and loss values as functions of learning epoch $N_{\rm epoch}$. The $F_{\rm 1}$-scores and loss values almost converge after the first few learning epochs. Since the $F_{\rm 1}$-scores (loss values) of the validation data are only slightly lower (higher) than those of the training data at later epochs of $N_{\rm epoch}\gtrsim15$, the 3D-CNN classifier shows no sign of significant overfitting. 

\begin{figure}
  \includegraphics[bb=0 0 1476 1398, width=\linewidth]{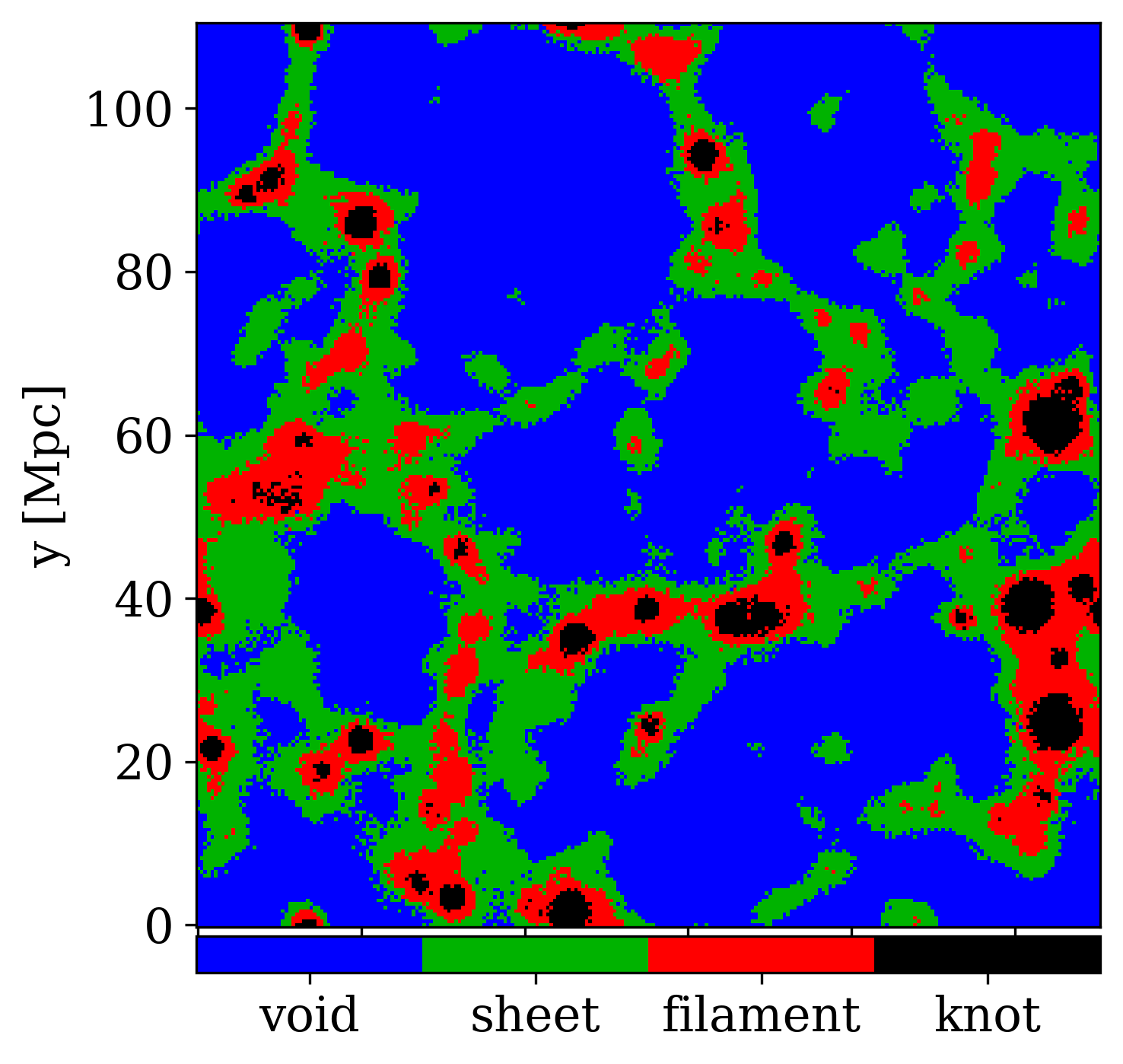}
\caption{Same as the bottom panel of Fig. \ref{DMmap} but plotting the labels predicted with the 3D-CNN model of the grid-based classification.}
\label{Grid-base_Pred}
\end{figure}
Fig. \ref{Grid-base_Pred} shows the map of predicted labels on the same slice of Fig. \ref{DMmap}. Unlike the distribution of the true labels, the boundary of the structures are blurred. This is probably because our cubic data have the side length of $10~{\rm Mpc}$, and this length limits the spatial resolution of the prediction. As we mention in Section \ref{CreateData}, however, decreasing the side length does not improve the accuracy of the prediction. Over-prediction of knot and filament regions are significant, and the knot regions are quite large in the prediction (see also Figs. \ref{CosWeb_GalBase} and \ref{CosWeb_Obs} in Section \ref{galbase}).

\begin{figure}
  \includegraphics[bb=0 0 769 524, width=\hsize]{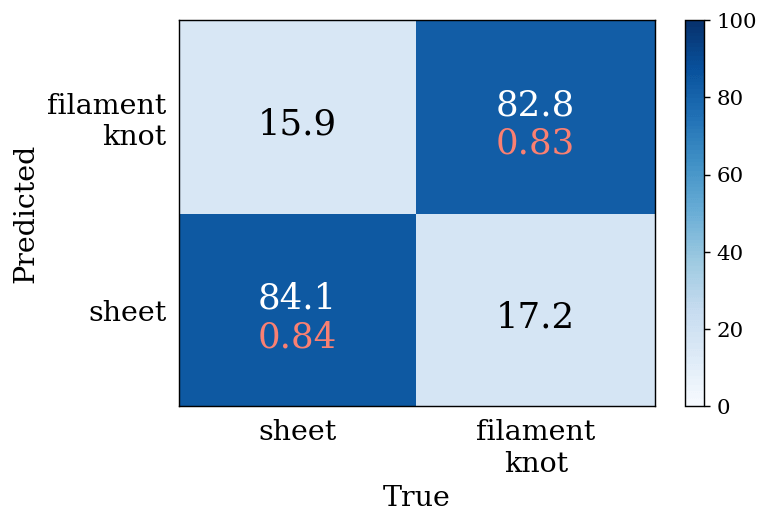}
\caption{The normalised confusion matrix of the grid-based classification into the two categories where voids are ignored and knots are merged with filaments. This binary classification is consistent with that in \citet{a:19}. The values in each cell indicate the same as in Fig. \ref{CM_grid_four}. The macro-average of the $F_{\rm 1}$-scores is $0.84$ for the test data.}
\label{CM_grid_two}
\end{figure}
\citet{a:19} has performed similar cosmic-structure classification with 3D-CNN on a U-Net architecture \citep[e.g.][]{UNet} although his deep-learning model learns DM density fields in large cosmological volumes and simultaneously classifies all spatial voxels therein. In terms of classifying spatial grid points, the situation of his study corresponds to the grid-based classification in ours. His study is, however, based on a binary classification where he ignores voids and merges knots with filaments. The largest difference from ours is that his model predicts the class labels from the DM density fields instead of galaxies. The performance of his model is $F_{\rm 1}=0.78$ and $0.72$ for sheets and filaments/knots. Fig. \ref{CM_grid_two} shows our result for the same binary classification where voids are ignored and knots are merged with filaments. Specifically, we randomly create 10000 data cubes centred on the sheet voxels and another set of 10000 cubes on the filament and knot voxels without distinguishing between the two. In our result shown in Fig. \ref{CM_grid_two}, the $F_{\rm 1}$-scores are $0.84$ and $0.83$ for sheets and filaments/knots, and the performance of our model is thus comparable with that of \citet{a:19}.\footnote{Although our simple model may be more accurate than that of \citet{a:19} in $F_{\rm 1}$-scores, the accuracy can depend on the resolutions of a simulation and the analysis of DM for the labelling. We use the tensors of velocity gradients (equation \ref{VelGraTendor}) to categorise the structures, whereas he does the second-order derivative of DM density fields.} This result demonstrates that galaxy distribution can be a substitution for DM density fields in predicting the cosmic structures with the models. It can, therefore, be feasible to classify the cosmic structures in the real Universe by using wide-field survey observations for galaxies such as SDSS. In addition, the above result proves that our method with the simple 3D-CNN classifier trained with the cubic data of galaxy distribution works as accurate as the U-Net architecture learning DM density fields in the previous study.

\subsection{Classifying galaxies with the grid-based model}
\label{galwithgridbase}
\begin{figure}
  \includegraphics[bb=0 0 769 493, width=\hsize]{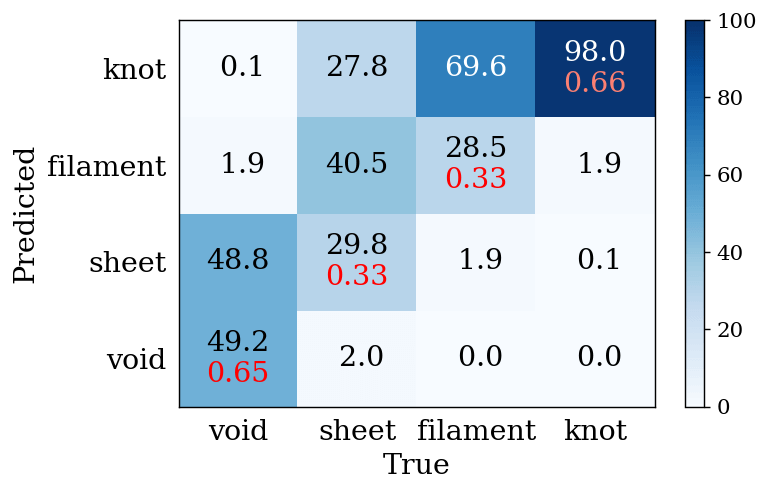}
\caption{Same as Fig. \ref{CM_grid_four} but classifying the galaxy-based cubic data with the model trained with the grid-based data in Fig. \ref{CM_grid_four}. The averaged $F_{\rm 1}$-score is $0.49$.}
\label{CM_galwithgrid_four}
\end{figure}
We stress, however, that the above models in Section \ref{gridbase} cannot classify galaxies since they are trained with the grid-based data. The grid-based classification uniformly samples spatial regions for each class label, whereas galaxies distribute with a bias towards high-density regions. For example, few galaxies are found in the middle of void regions, and most of the galaxies labelled as voids reside around boundaries of void regions which are close to sheets.  Accordingly, in the galaxy-based classification where the data cubes are centred on galaxies, the cubic regions centred on the void galaxies are generally similar to sheet regions. Sheet and filament galaxies are also subject to the same bias since most of them are found in regions close to filament and knot regions, respectively. To demonstrate the influence of this bias, we classify the galaxy-based data cubes with the model trained for the grid-based classification that is used for Fig. \ref{CM_grid_four}. The galaxy-based data cubes are centred on galaxies selected randomly, and we create a set of 10000 data for each label. Fig. \ref{CM_galwithgrid_four} shows the result, where the predictions are significantly biased towards contiguous classes with higher densities except for knots. In comparison with Fig. \ref{CM_grid_four}, the $F_{\rm 1}$-scores are significantly lower in Fig. \ref{CM_galwithgrid_four}. For knot galaxies, although the true positive fraction is quite high with $98.0$ per cent, the false positive predictions to knots from sheets and filaments also largely increase to $27.8$ and $69.6$ per cent. As a result, the $F_{\rm 1}$-score of knots becomes lower as well as the other classes. The macro-averaged $F_{\rm 1}$-score among the four classes is $0.49$. Thus, we cannot adopt the model trained with the grid-based classification to galaxies. This implies that it is difficult to classify spatial regions close to boundaries between contiguous classes.  

The above models trained with the grid-based data can, however, be useful for other purposes such as classifying spatial regions. For example, it can define extents of void regions and lengths of filament structures in the real Universe. Especially, the size of `the super void' and statistics of the filament lengths are used as tests of the models of cosmology \citep[e.g.][]{ik:06,ik:07,spc:08}. A point worthy to mention here is that our methods use observable galaxies to classify the cosmic structures, but the class labels are obtained from velocity gradients of unobservable DM (equation \ref{VelGraTendor}). This is qualitatively different from the previous classification methods described in Section \ref{Intro}.

\subsection{The galaxy-based classification}
\label{galbase}
\subsubsection{Without observational restrictions}
\label{noobsrest}
As we show in Fig. \ref{CM_galwithgrid_four}, the grid-based model cannot classify galaxies. To classify galaxies, we need to train the 3D-CNN models with the galaxy-based data whose classification points are on galaxies. We here ignore observational restrictions, include all the haloes having stars in the simulation and use their actual positions in creating the learning data.

\begin{figure}
  \includegraphics[bb=0 0 769 493, width=\linewidth]{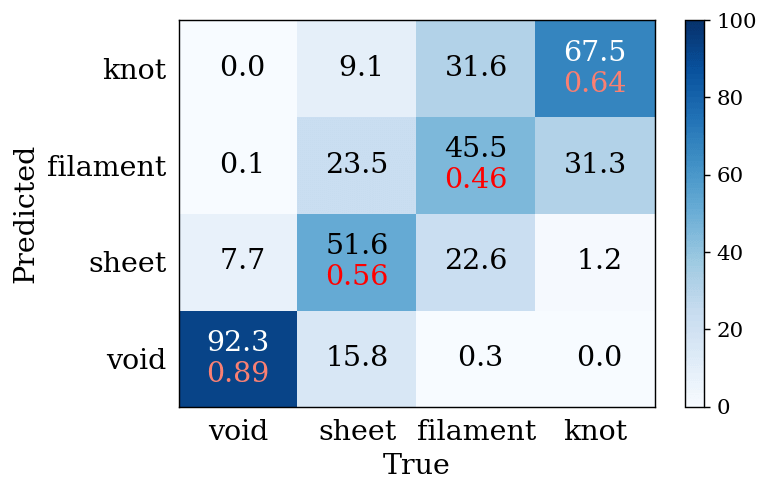}
  \includegraphics[bb=0 0 769 524, width=\linewidth]{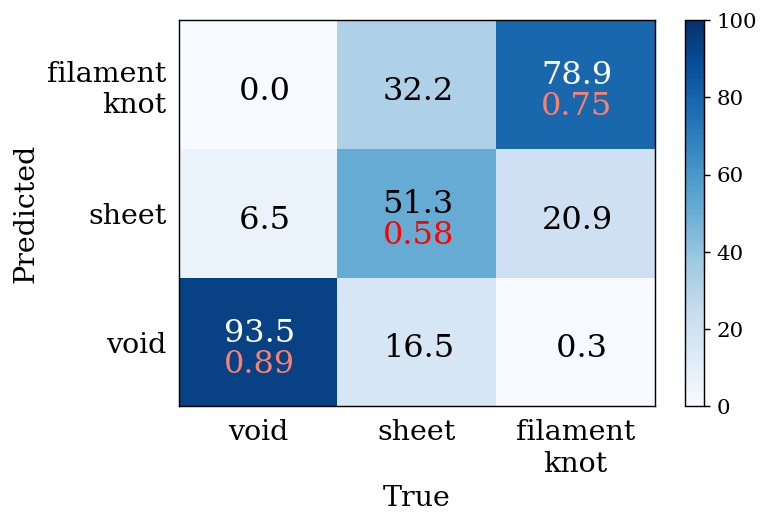}
  \includegraphics[bb=0 0 769 555, width=\linewidth]{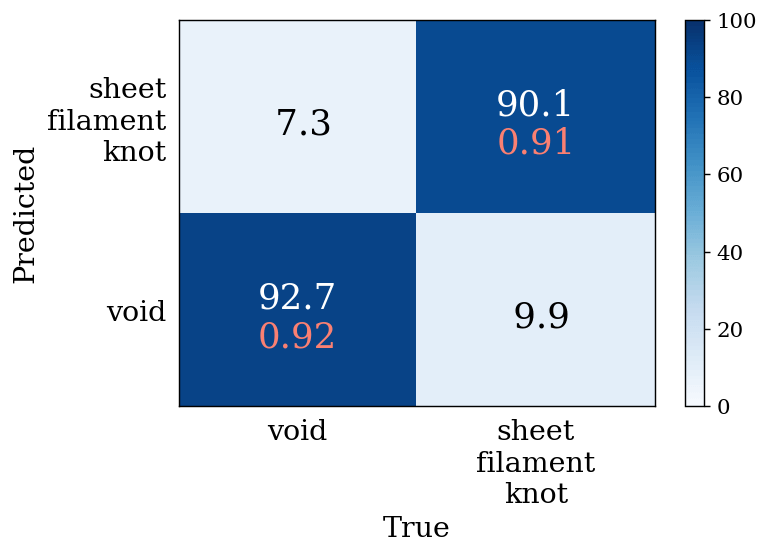}
\caption{\textit{Top}: same as \ref{CM_grid_four} but the galaxy-based classification into the four categories. \textit{Middle}: ternary classification where knots are merged with filaments. \textit{Bottom}: binary classification where knots, filaments and sheets are merged into a single class. The macro-averages of $F_{\rm 1}$-scores among the classes are $0.64$, $0.74$ and $0.92$ in the top, middle and bottom panels, whereas the `number-averaged' $F_{\rm 1}$-scores are $0.58$, $0.68$ and $0.91$.}
\label{CM_galbase_all}
\end{figure}
\begin{figure}
  \includegraphics[bb=0 0 730 500, width=\hsize]{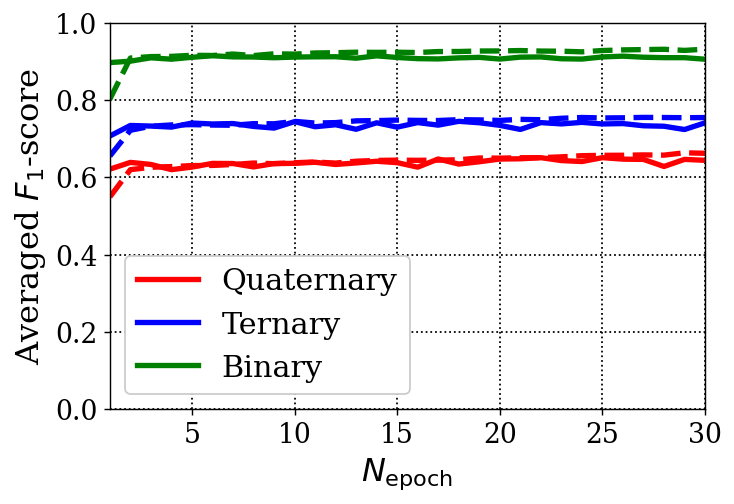}
\caption{The macro-averages of $F_{\rm 1}$-scores for the validation data as functions of epoch $N_{\rm epoch}$ in the learning processes of the quaternary, ternary and binary classification in Fig. \ref{CM_galbase_all}.}
\label{LA_galbase_all}
\end{figure}
The top panel of Fig. \ref{CM_galbase_all} shows the normalised confusion matrix and $F_{\rm 1}$-scores in the case of the quaternary classification. In comparing with the grid-based classification (Fig. \ref{CM_grid_four}), the $F_{\rm 1}$-scores become lower except for voids. The macro-average among the four classes is $F_{\rm 1}=0.64$ and decreases by $0.1$ from the grid-based classification. This implies that it is more difficult to classify galaxies than the spatial grid points due to the bias we argue in Section \ref{gridbase}. It should be noted, however, that the classification of sheets and filaments is relatively inaccurate, and the galaxies in the two classes account for $79$ per cent of all galaxies (see Section \ref{sim_csc}). Accordingly, the total performance for classifying all galaxies is dominated by the accuracy of sheets and filaments. Hence, if we evaluate the average of the  $F_{\rm 1}$-scores weighted by the numbers of galaxies in the four classes, the `number-averaged' $F_{\rm 1}$-score is $0.58$. Although we have sampled 10000 cubic data for each labels in this study, this sample size may be insufficient for the large classes such as voids in the grid-based classification and sheets and filaments in the galaxy-based classification. If it is the case, our model can become less robust. In Appendix \ref{inbalance}, we examine the effect by repeating the same computations and find the fluctuation of $F_{\rm 1}$-scores to be small.

In the middle panel of Fig. \ref{CM_galbase_all}, we merge the knots with filaments by sampling 10000 galaxies from the two classes without distinguishing them. In this ternary classification, the macro-averaged $F_{\rm 1}$-score increases to $0.74$. However, merging the knots with filaments does not appear to improve the accuracy of the voids or sheets, and the $F_{\rm 1}=0.58$ for the sheets means that it is still difficult to classify the sheets accurately. The number-averaged $F_{\rm 1}$-score is $0.68$. However, since the class of filaments/knots is relatively accurate with $F_{\rm 1}=0.75$, this model could be used to identify galaxies in cosmic streams connecting knot regions. Finally, in the bottom panel of Fig. \ref{CM_galbase_all}, we merge the knots and filaments with the sheets. The macro-averaged $F_{\rm 1}$-score is $0.92$ in this binary classification, and this model can identify void galaxies with the high accuracy. Fig. \ref{LA_galbase_all} indicates the macro-averaged $F_{\rm 1}$-scores in the three cases of Fig. \ref{CM_galbase_all} as functions of learning epoch $N_{\rm epoch}$. In the galaxy-based classification, we find no significant overfittings until the last epoch, and the macro-averaged $F_{\rm 1}$-scores converge after the first few epochs.

\begin{figure*}
  \includegraphics[bb=0 0 4832 2370, width=\hsize]{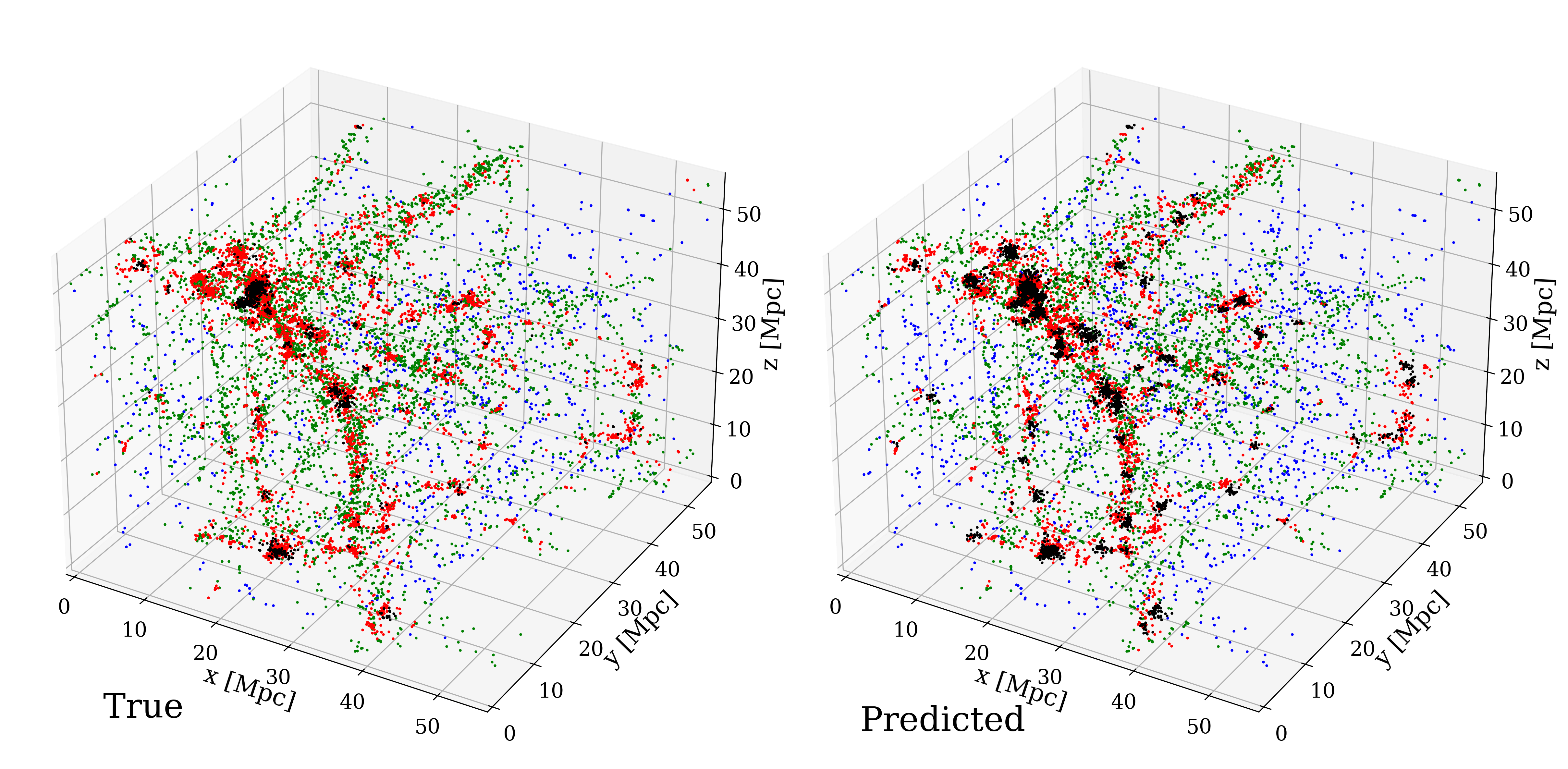}
\caption{Spatial distribution of galaxies in a one-eighth volume of the simulation. The galaxies are coloured with true (left) and predicted (right) labels: void, sheet, filament and knot galaxies with blue, green, red and black. We here use the model for the galaxy-based classification without observational restriction (the top panel of Fig. \ref{CM_galbase_all}), plot galaxies with stellar masses of $M_{\rm star}>10^{9.5}~{\rm M_\odot}$ for visibility.}
\label{CosWeb_GalBase}
\end{figure*}
\begin{figure*}
  \begin{minipage}[b]{0.247\linewidth}
    \centering
  \includegraphics[bb=0 0 2370 4747, width=\linewidth]{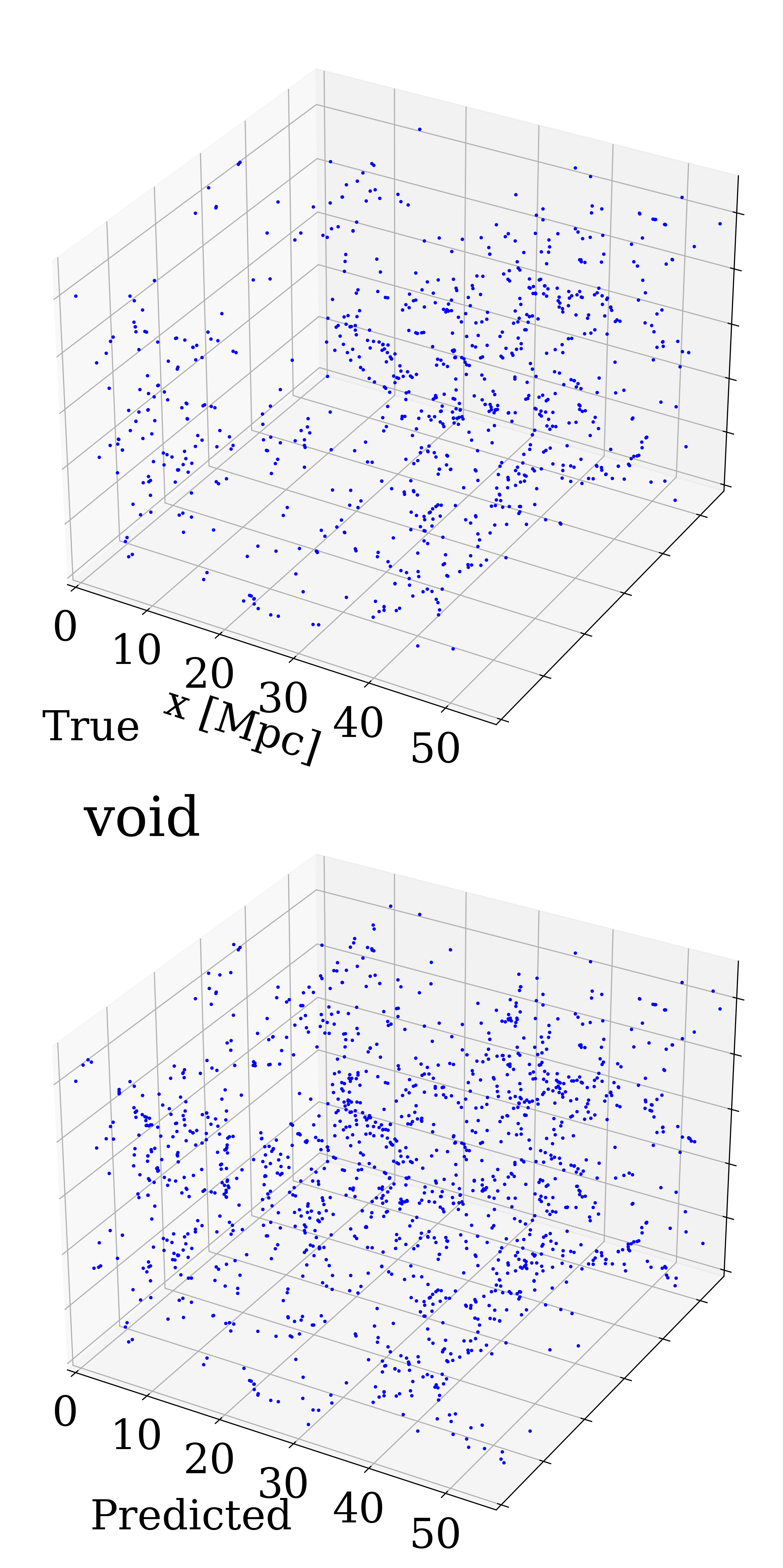}
  \end{minipage}
  \begin{minipage}[b]{0.247\linewidth}
    \centering
  \includegraphics[bb=0 0 2370 4747, width=\linewidth]{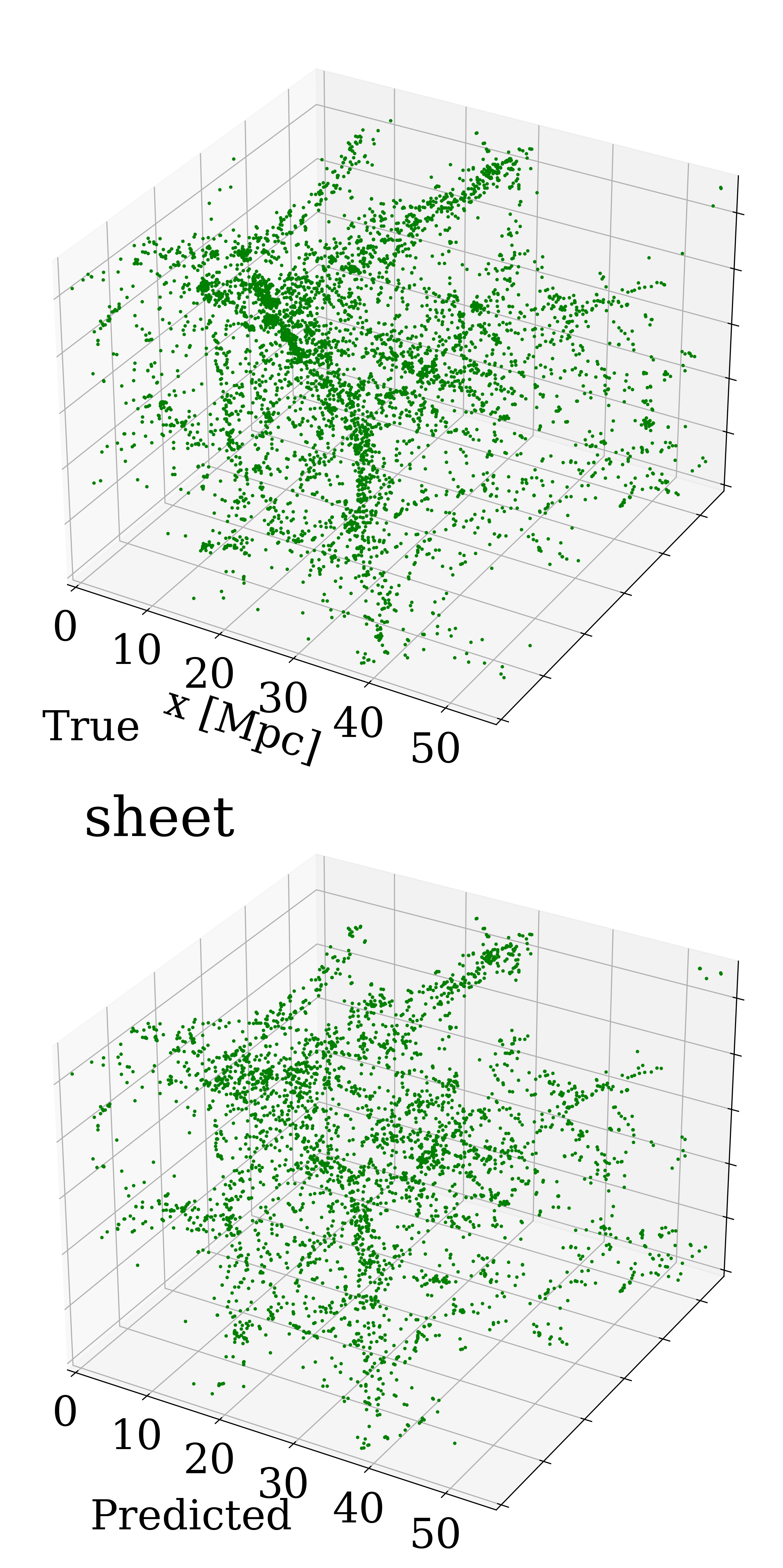}
  \end{minipage}
  \begin{minipage}[b]{0.247\linewidth}
    \centering
  \includegraphics[bb=0 0 2370 4747, width=\linewidth]{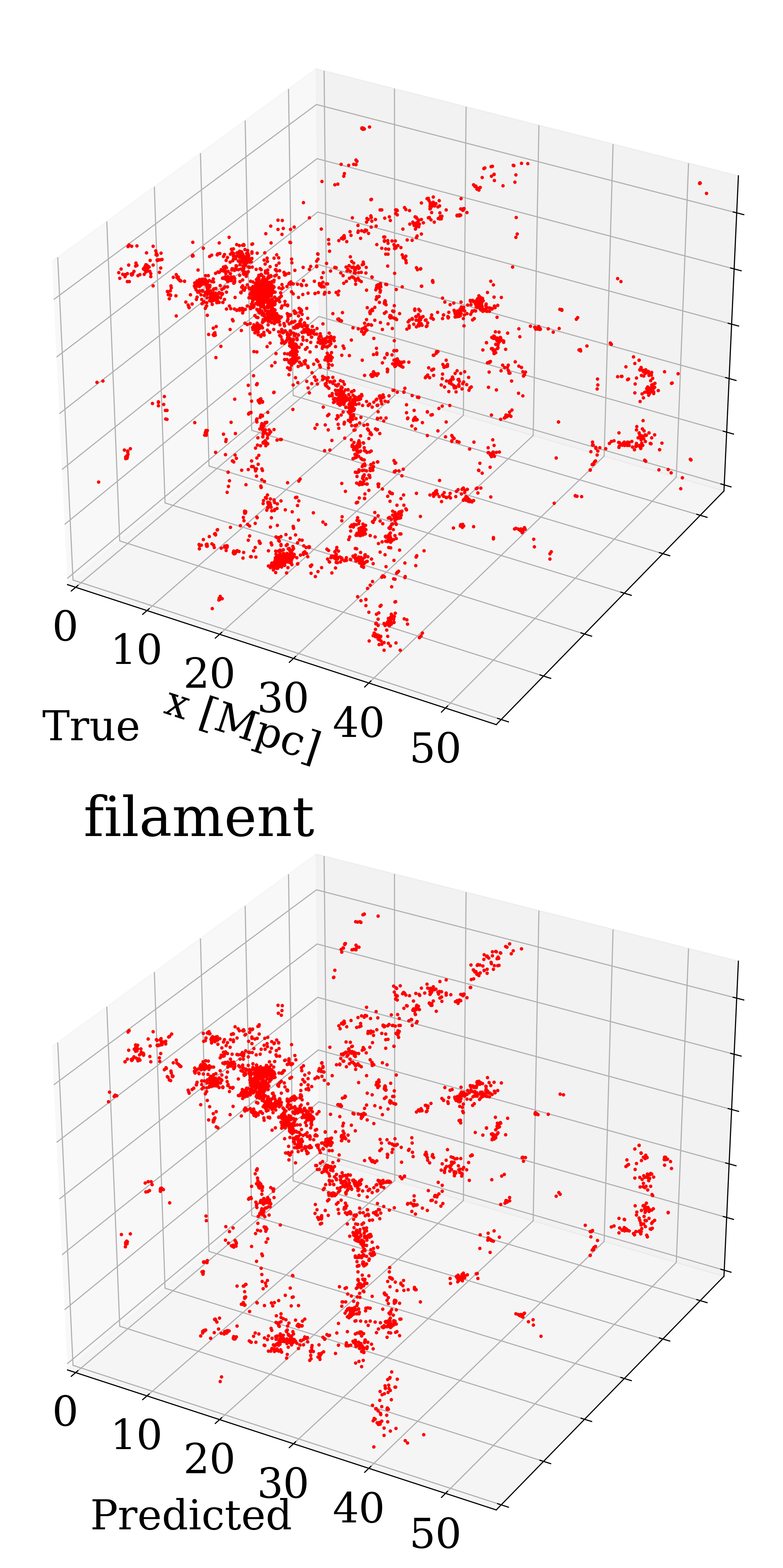}
  \end{minipage}
  \begin{minipage}[b]{0.247\linewidth}
    \centering
  \includegraphics[bb=0 0 2370 4747, width=\linewidth]{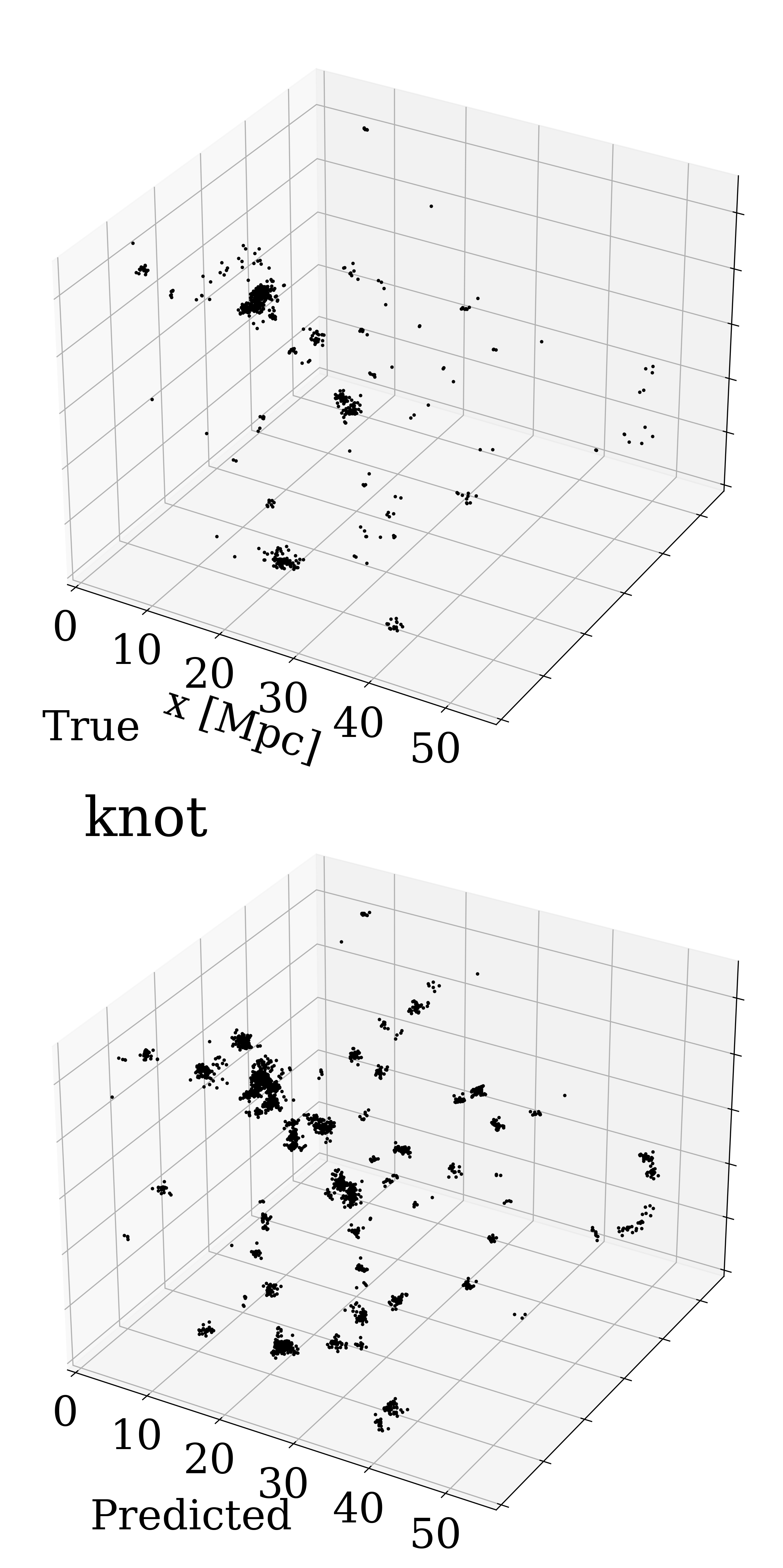}
  \end{minipage}
\caption{Same as Fig. \ref{CosWeb_GalBase} but plotting the galaxies separately into the four categories for the true (top) and predicted (bottom) labels.}
\label{CosWeb_GalBase_Each}
\end{figure*}
To compare the true and predicted labels in a three-dimensional space, we extract a one-eighth volume with a side length of $55~{\rm Mpc}$ from the simulation and use the galaxies therein to create new test data. The other galaxies in the remaining volume are used to create new sets of 10000 data cubes for each of the four classes, and we train the model with the new data in the same manner. We confirm that the validation accuracy and $F_{\rm 1}$-scores are similar to the above case in Fig. \ref{CM_galbase_all}. In Fig. \ref{CosWeb_GalBase}, we plot the positions of galaxies in the one-eighth volume, where they are coloured with their true and predicted labels in the left and right panels. Fig. \ref{CosWeb_GalBase_Each} shows the same but plots the galaxies separately for each label. The most conspicuous error would be the over-prediction of the number of knot galaxies in the right panel. From the top panel of Fig. \ref{CM_galbase_all}, nearly $30$ per cent of filament galaxies are erroneously classified as knots in the quaternary classification. Note that the filament and knot galaxies are the second largest and the smallest populations (see Section \ref{sim_csc}), and the number of filaments is nearly five times larger than that of knot galaxies in the whole simulation. Therefore, the majority of galaxies predicted as knots are erroneous classification of (true) filament galaxies. However, the galaxies predicted as knots/filaments appear to delineate the filamentary structures where most of the (true) filament galaxies reside. Similar contamination is expected between the sheet and void galaxies. The number of sheet galaxies in the true labels is nearly three times larger than that of void galaxies, and $\sim15$ per cent of the sheet galaxies are erroneously classified as voids although the accuracy for the voids is high. Accordingly, nearly one-third of the galaxies predicted as voids stem from the erroneous classification of sheets.

\subsubsection{With observational restrictions}
\label{withobsrest}
To apply our 3D-CNN models to actual observations, we need to take observational restrictions into consideration such as limiting magnitudes and errors on distance measurements. Because our models use three-dimensional distribution of galaxies, spectroscopic determinations of distances (redshifts $z$) are required for observed galaxies. Generally, a limiting magnitude of spectroscopy is more severe than that of photometry. We cannot include galaxies fainter than the spectroscopic limiting magnitude in our samples since they lack distance measurements. In the SDSS observations, the limit on apparent magnitude is estimated to be $m_r=17.75~{\rm mag}$ in $r$-band (see Appendix \ref{SDSSlimit}), which corresponds to the absolute magnitude of $M_r=-17.25~{\rm mag}$ at a distance of $100~{\rm Mpc}$ from the Earth. If we assume that all of the simulated galaxies are at $100~{\rm Mpc}$, the numbers of galaxies in our samples reduce to $1051$, $9251$, $5525$ and $1413$ for voids, sheets, filaments and knots, respectively. The red histogram in Fig. \ref{HOF} shows the number of the `observable' galaxies brighter than $M_r=-17.25~{\rm mag}$, and most of the galaxies more massive than $\sim10^{9.5}~{\rm M_\odot}$ in DM mass are captured in this sample selection. Assuming that distance measurements are available for all galaxies brighter than the limiting magnitude, we use all the `observable' galaxies in the simulation.

Although the above assumption for the distance is arbitrary, we expect that the distance of $100~{\rm Mpc}$ is within a range where our models can work efficiently for the SDSS catalogue (see Section \ref{appSDSS}). If we assume a closer distance $\ll100~{\rm Mpc}$, the survey area of SDSS becomes too small to capture the variety of the cosmic structures, especially the dense environments such as filament and knot. We find that the number of observed galaxies largely increases at $\simeq100~{\rm Mpc}$ in the SDSS catalogue. If we assume a larger distance $\gg100~{\rm Mpc}$, on the other hand, the limit on apparent magnitude becomes too severe to obtain sufficiently large samples of the simulated galaxies for the training data. Especially, only a handful of void galaxies are brighter than the magnitude limit at a distance $\gg100~{\rm Mpc}$. Our 3D-CNN models can be less robust in such a case.\footnote{To apply our 3D-CNN models to more distant galaxies, we need to utilise other cosmological simulations resolving larger volumes such as TNG300 (see also Appendix \ref{TNG50} for the case of TNG50-1).}

In addition, line-of-sight positions (i.e. distances) of galaxies cannot be directly measured but are estimated from their recession velocities. Therefore, assuming uniform cosmic expansion in the nearby Universe, distance measurements of galaxies are affected by line-of-sight components of their peculiar velocities, $v_{\rm los}$, as
\begin{equation}
    x_{\rm obs}=x_{\rm true} + \frac{v_{\rm los}}{H_0},
    \label{losv}
\end{equation}
where $x_{\rm obs}$ and $x_{\rm true}$ are estimated and true lines-of-sight positions of a galaxy. This effect can be significant in knot regions since peculiar velocities are generally large in galaxy clusters; the clusters are stretched along the line-of-sight directions, and this effect is know as the `fingers-of-god effect'. Because the assumed distance of $100~{\rm Mpc}$ is significantly larger than the size of the cubic region for our training data, $10~{\rm Mpc}$, we can assume the lines of sight to be parallel for galaxies in the cubic region.

\begin{figure}
  \includegraphics[bb=0 0 769 493, width=\linewidth]{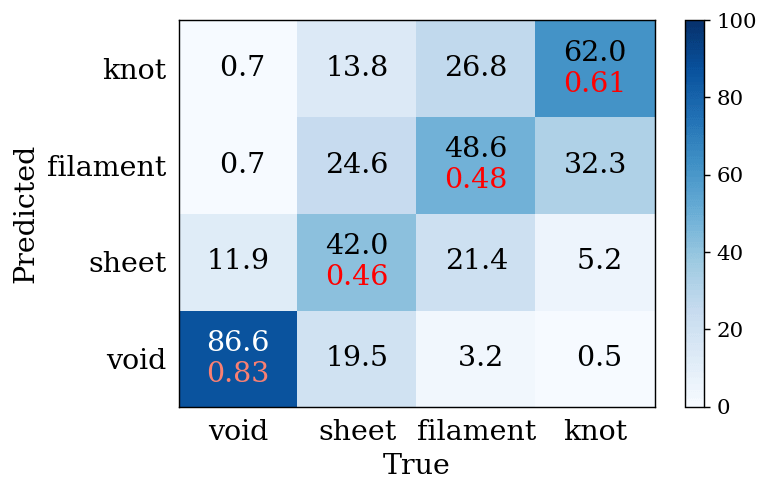}
  \includegraphics[bb=0 0 769 524, width=\linewidth]{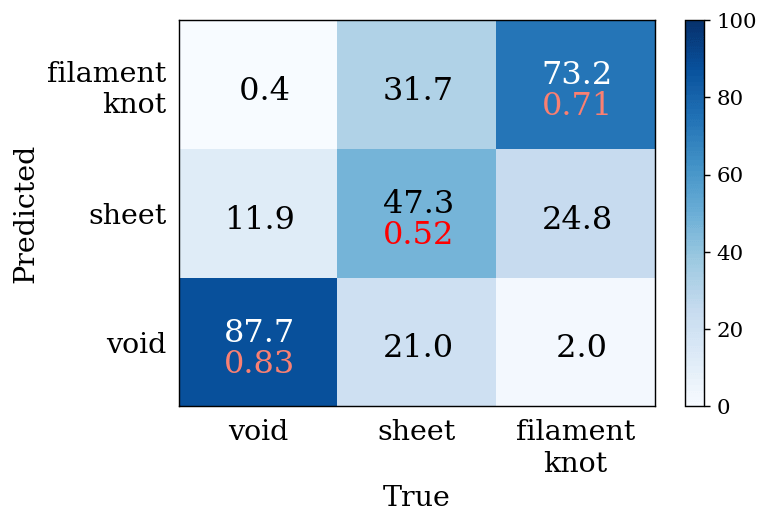}
  \includegraphics[bb=0 0 769 555, width=\linewidth]{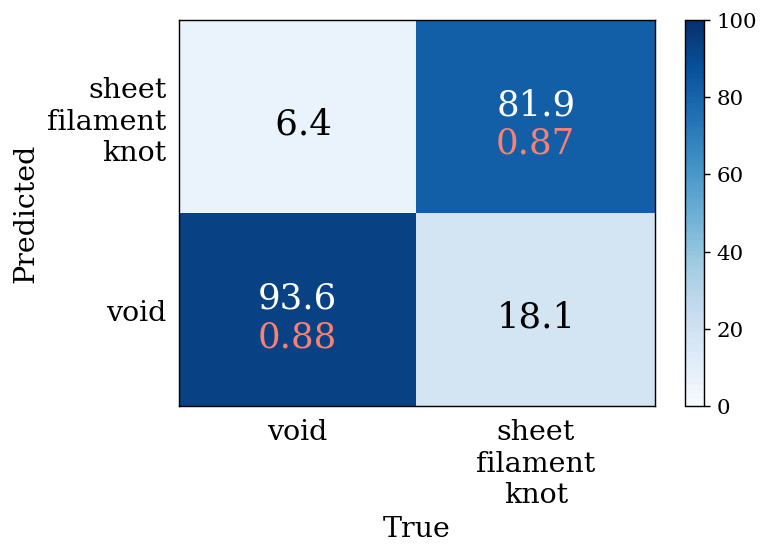}
\caption{Same as Fig. \ref{CM_galbase_all} but excluding galaxies fainter than $M_r=-17.25~{\rm mag}$ and including the effect of $v_{\rm los}$ in creating the data. The macro-averaged $F_{\rm 1}$-scores are $0.60$, $0.68$ and $0.88$ for the test data in the top, middle and bottom panels, whereas the number-averaged $F_{\rm 1}$-scores are $0.5$, $0.62$ and $0.87$.}
\label{CM_galbaseObs100Mpc_all}
\end{figure}
\begin{figure}
  \includegraphics[bb=0 0 730 500, width=\hsize]{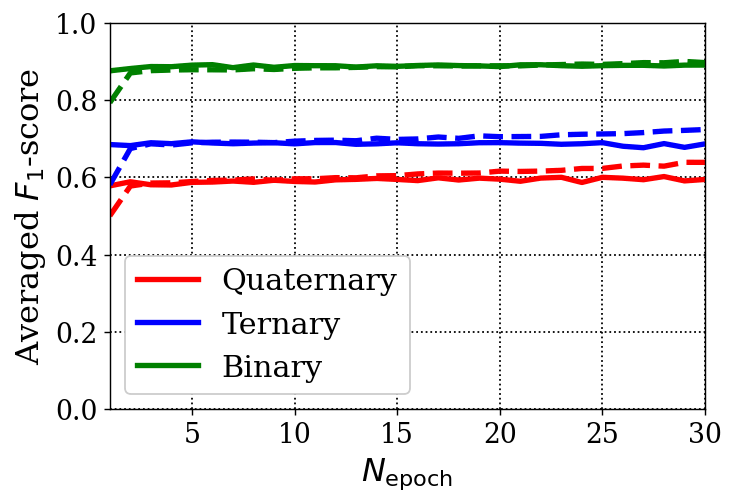}
\caption{Same as Fig. \ref{LA_galbase_all} but for the models used in Fig. \ref{CM_galbaseObs100Mpc_all}.}
\label{LA_galbaseObs100Mpc_all}
\end{figure}
Taking into account the above observational restrictions, we again build the models with the galaxy-based classification. First, we exclude galaxies fainter than the limiting magnitude and randomly select a galaxy as a classification point. After we randomly rotate galaxies around the classification point and set a line of sight, we shift lines-of-sight positions of all galaxies according to equation (\ref{losv}). A data cube is centred on the classification point affected by its $v_{\rm los}$ while it keeps its class label assigned at the original position. We sample 10000 galaxies with replacement from the entire simulation for each label and create the cubic data to train our models. Fig. \ref{CM_galbaseObs100Mpc_all} shows the normalised confusion matrixes and $F_{\rm 1}$-scores for the quaternary (top), ternary (middle) and binary (bottom) classification. In the three cases, the averaged $F_{\rm 1}$-scores decrease from those in Fig. \ref{CM_galbase_all} by $0.04$--$0.06$. However, the overall trends hardly change; the classification of sheets and filaments is relatively inaccurate with the low $F_{\rm 1}$-scores, and void galaxies are identified the most accurately. Note again that the sheets and filaments are the dominant classes in number. The number-averaged $F_{\rm 1}$-scores described in Section \ref{noobsrest} are $0.5$, $0.62$ and $0.87$ in the quaternary, ternary and binary classification. Fig. \ref{LA_galbaseObs100Mpc_all} indicates the macro-averaged $F_{\rm 1}$-scores for the validation data as functions of learning epoch $N_{\rm epoch}$. The results appear to be similar to Fig. \ref{LA_galbase_all} although there may be hints of weak overfittings at $N_{\rm epoch}\gtrsim25$ in the quaternary and ternary classification.

\begin{figure*}
  \includegraphics[bb=0 0 4832 2370, width=\hsize]{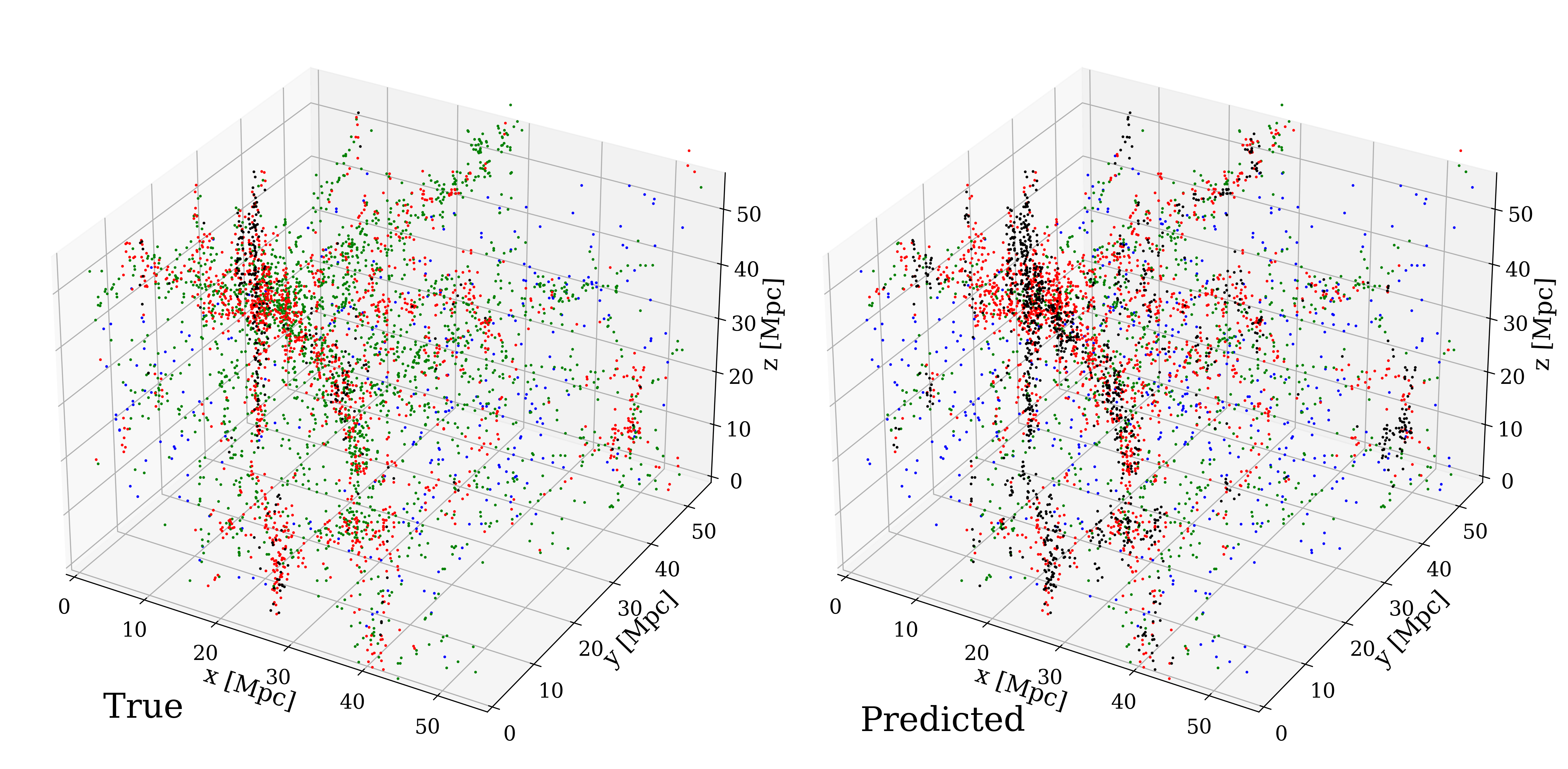}
\caption{Same as \ref{CosWeb_GalBase} but excluding galaxies fainter than $M_r=-17.25~{\rm mag}$ and including the effect of $v_{\rm los}$. The line of sight is along the $z$-axis. We here plot all the galaxies brighter than the limiting magnitude.}
\label{CosWeb_Obs}
\end{figure*}

\begin{figure*}
  \begin{minipage}[b]{0.247\linewidth}
    \centering
  \includegraphics[bb=0 0 2370 4747, width=\linewidth]{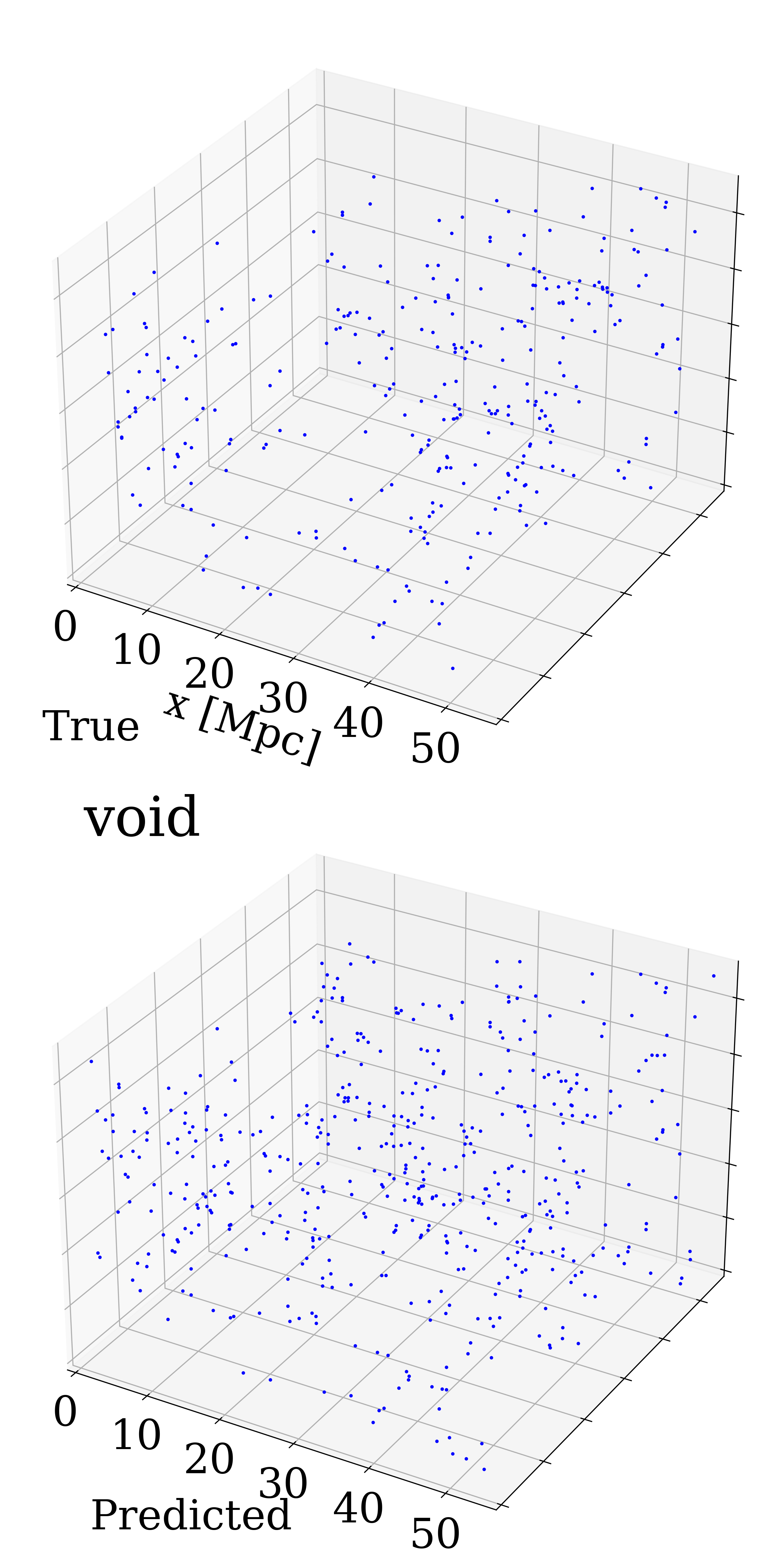}
  \end{minipage}
  \begin{minipage}[b]{0.247\linewidth}
    \centering
  \includegraphics[bb=0 0 2370 4747, width=\linewidth]{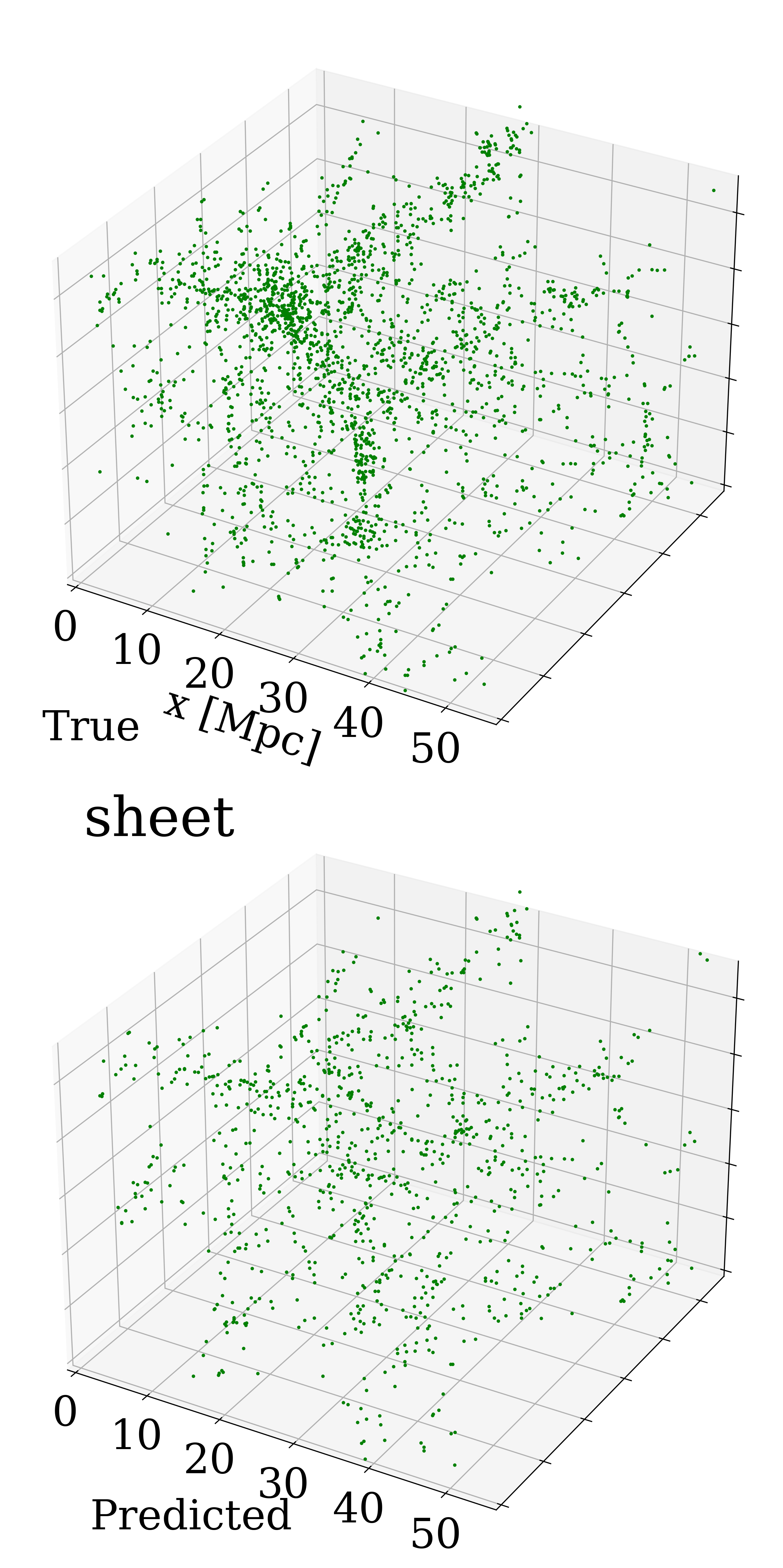}
  \end{minipage}
  \begin{minipage}[b]{0.247\linewidth}
    \centering
  \includegraphics[bb=0 0 2370 4747, width=\linewidth]{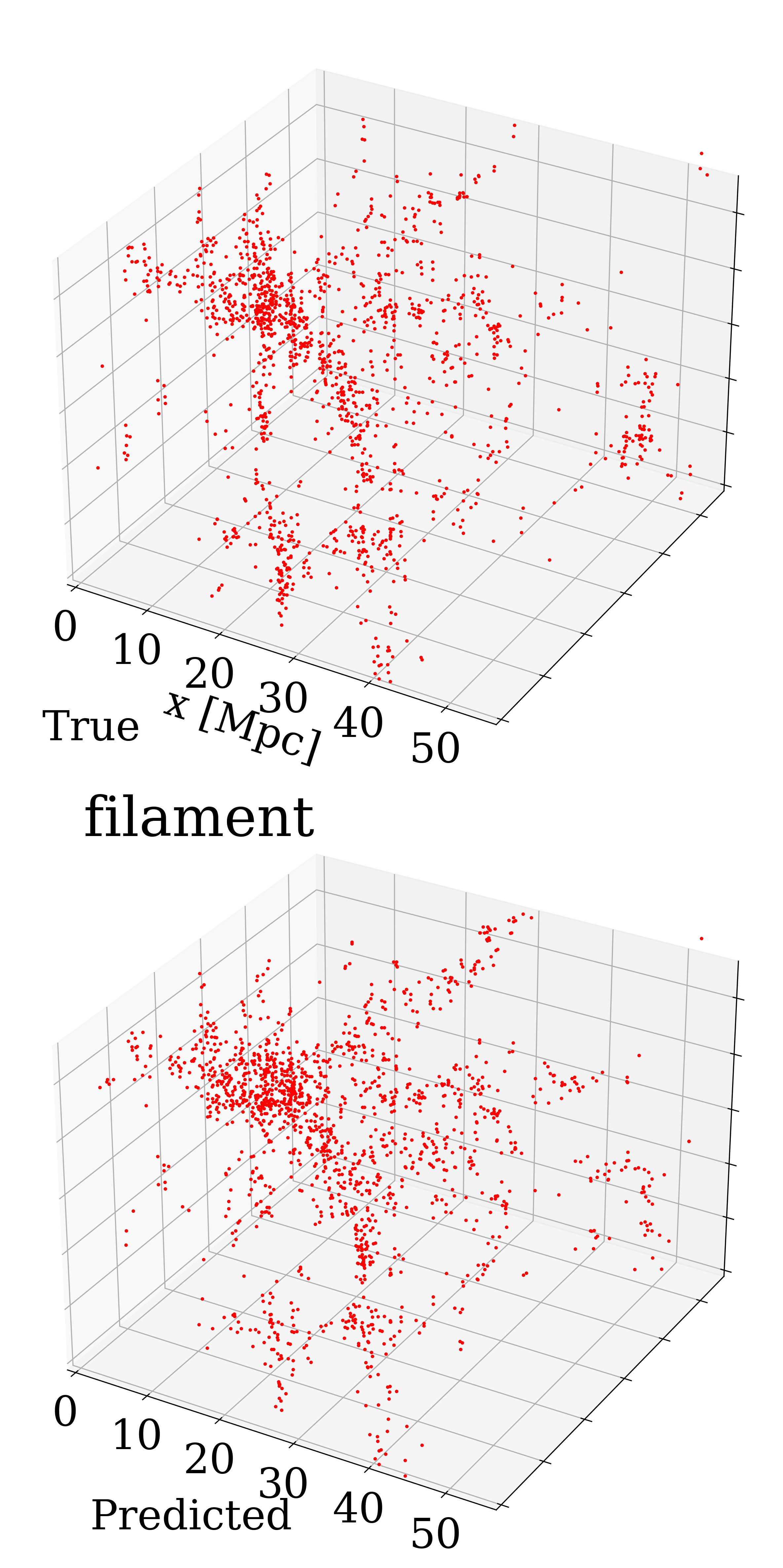}
  \end{minipage}
  \begin{minipage}[b]{0.247\linewidth}
    \centering
  \includegraphics[bb=0 0 2370 4747, width=\linewidth]{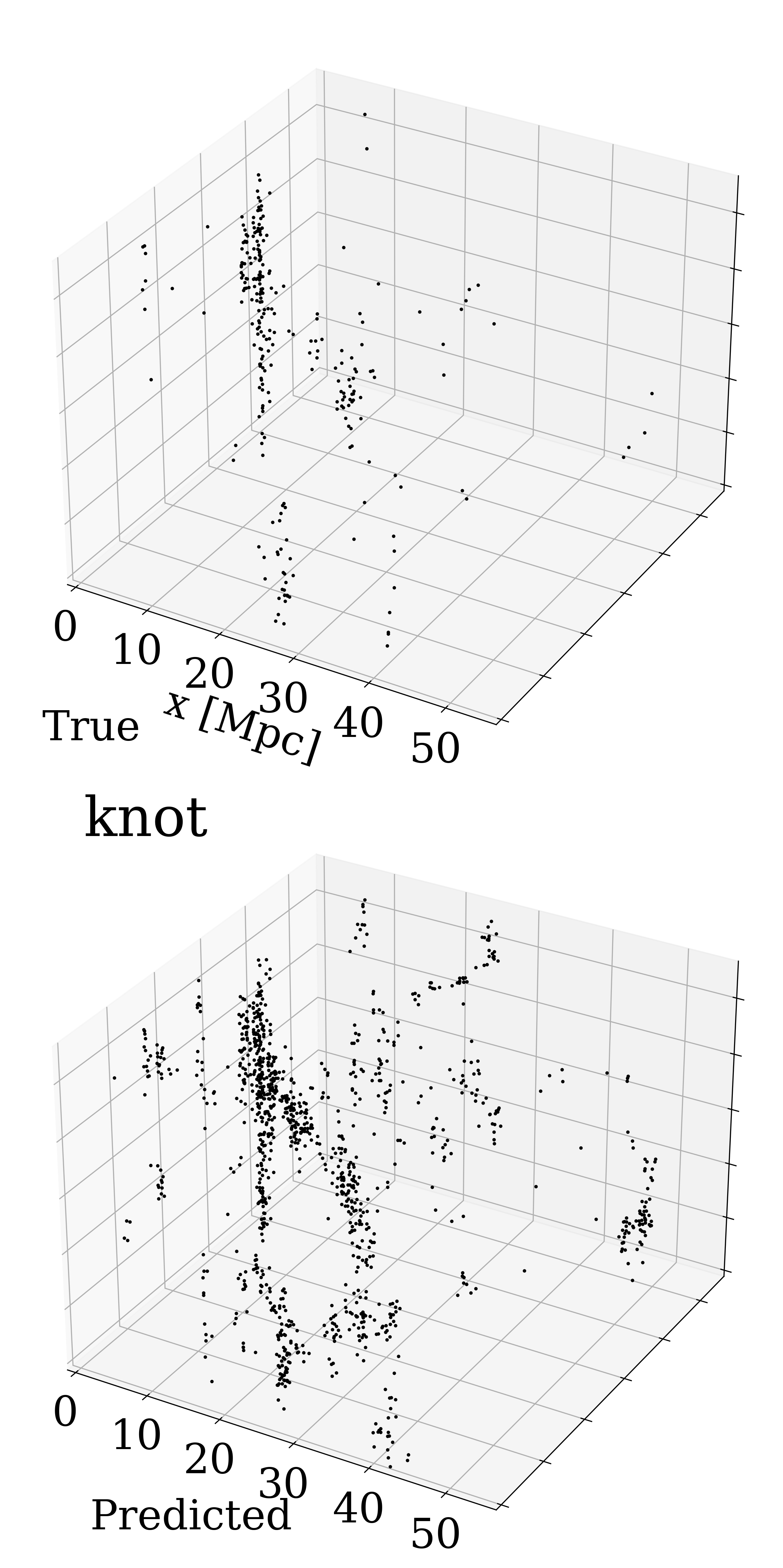}
  \end{minipage}
\caption{Same as Fig. \ref{CosWeb_Obs} but plotting the galaxies separately into the four categories in the true (top) and predicted (bottom) labels.}
\label{CosWeb_Obs_Each}
\end{figure*}

As is done for Figs. \ref{CosWeb_GalBase} and \ref{CosWeb_GalBase_Each}, we extract the same one-eighth volume in the simulation and create new test data, and the rest is used for training and validation data. Fig. \ref{CosWeb_Obs} shows the comparison between the true and predicted class labels in the three-dimensional space, where the line of sight is pointed along the $z$-axis. Fig. \ref{CosWeb_Obs_Each} shows the same but plots the galaxies separately for each label. Since the plotted volumes in Figs. \ref{CosWeb_Obs} and \ref{CosWeb_Obs_Each} are the same as in Figs. \ref{CosWeb_GalBase} and \ref{CosWeb_GalBase_Each}, the comparison between them shows the influence of the observational restrictions. Because of the limiting magnitude, there are a smaller number of galaxies available for the data. Dense regions corresponding to galaxy clusters are significantly elongated along the line of sight ($z$-axis) due to the effect of $v_{\rm los}$ (equation \ref{losv}), and the distribution of the (true) knot and filament galaxies is especially affected (the top panels of Fig. \ref{CosWeb_Obs_Each}). In comparing the true and predicted labels in Figs. \ref{CosWeb_Obs} and \ref{CosWeb_Obs_Each}, the significant over-prediction of knot galaxies is seen as in Figs. \ref{CosWeb_GalBase} and \ref{CosWeb_GalBase_Each}. Among the galaxies brighter than the magnitude limit, $54$ per cent are (true) sheets, and the number of these sheet galaxies is $8.8$ and $1.7$ times larger than those of the void and filament galaxies, respectively. In the prediction, nearly $16$ and $25$ per cent of the sheet galaxies are mistaken for voids and filaments, respectively (the top panel of Fig. \ref{CM_galbaseObs100Mpc_all}). Therefore, the erroneous classification of the sheets is also significant for the voids and filaments in the predicted labels of Figs. \ref{CosWeb_Obs} and \ref{CosWeb_Obs_Each}. However, the galaxies predicted as knots/filaments are mainly found along the filamentary structures and in the cluster regions elongated by the effect of $v_{\rm los}$ (see also Section \ref{improve}). On the other hand, the galaxies predicted as sheets appear to distribute more diffusely than the true sheet galaxies (Fig. \ref{CosWeb_Obs_Each}).


\subsection{Application to observations}
\label{appSDSS}
Our models taking into account the observational restrictions can be directly applied to observations. Using the models obtained for Fig. \ref{CM_galbaseObs100Mpc_all}, we here classify observed galaxies. From the galaxy catalogue of SDSS DR12 for the north side ($100\lesssim$ RA $\lesssim300~{\rm degree}$), we extract galaxies and quasars\footnote{Hereafter, we do not distinguish quasars from galaxies in the SDSS data.} that have no warning flags for their redshift measurements, and convert their redshifts $z$ to distances $d_{\rm obs}$ by assuming the uniform expansion of the local Universe. We do not take into account uncertainties of the redshift measurements since these are typically $\Delta z/z\sim10^{-4}$. We use galaxies within a range of $d_{\rm obs}=85$--$115~{\rm Mpc}$ and create data cubes whose classification points are centred on all the galaxies within $d_{\rm obs}=90$--$110~{\rm Mpc}$. In this narrow distance range, the limits on absolute magnitude are nearly constant with the small variation from $M_r=-17.02$ to $-17.46~{\rm mag}$. We can, therefore, consider the SDSS sample to be approximately volume-limited, and we use the models built by assuming the simulated galaxies to be at $100~{\rm Mpc}$ in Section \ref{withobsrest}. Our models can, however, classify galaxies at other distances by changing the absolute magnitude limit $M_r$ depending on the distances and retraining the models as long as the distance is not too large. We do not classify galaxies close to the boundaries of the survey area since our models cannot treat such galaxies if their data cubes overlap with the boundaries.

\begin{figure*}
  \includegraphics[bb=0 0 3187 1429, width=\hsize]{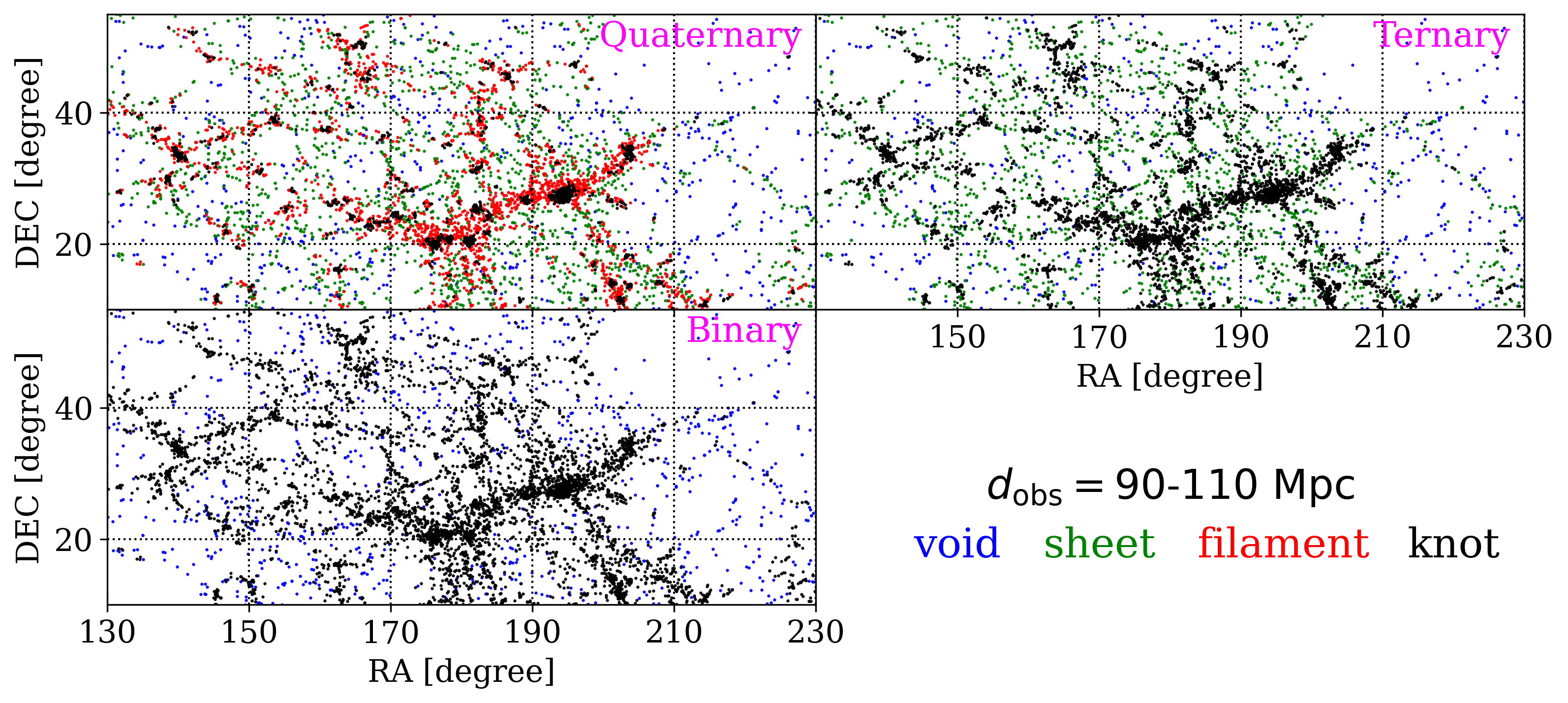}
\caption{Classification of the SDSS galaxies in the distance range from $d_{\rm obs}=90$ to $110~{\rm Mpc}$ using the models obtained in Fig. \ref{CM_galbaseObs100Mpc_all}. We include all galaxies with spectroscopic redshift measurements in the region. In the ternary (right) and binary (bottom left) classification, galaxies of the merged classes are plotted with black.}
\label{SDSS_classification}
\end{figure*}
Fig. \ref{SDSS_classification} shows the predicted class labels for the SDSS galaxies within $d_{\rm obs}=90$--$110~{\rm Mpc}$. In the quaternary classification (top left), we find $628$ ($9.1$), $1895$ ($27.6$), $2613$ ($38.1$) and $1730$ ($25.2$) galaxies (per cent) in the classes of void, sheet, filament and knot in the plotted region (RA$=130$--$230^\circ$ and DEC$=10$--$55^\circ$). The galaxies classified as filaments distribute along the structures resembling a web studded with compact groups of the knot galaxies. The sheet galaxies are found around the filaments, and the voids reside in the diffuse inter-filament regions. These features appear to agree with the intuitive recognition of the cosmic structures. Meanwhile, it should be reminded that the models cannot classify sheet and filament galaxies very accurately as shown in the top panel of Fig. \ref{CM_galbaseObs100Mpc_all}, and the $F_{\rm 1}$-scores are $0.46$ and $0.48$ for sheet and filament galaxies in the quaternary classification. From the analogy of Figs. \ref{CM_galbaseObs100Mpc_all}, \ref{CosWeb_Obs} and \ref{CosWeb_Obs_Each}, a significant fraction of the galaxies identified as knots can be misclassification of (true) sheets.

In the cases of the ternary (the right panel in Fig. \ref{SDSS_classification}) and binary (bottom left) classification, the merged classes are plotted in black. In the ternary classification, $548$ ($8.0$), $1788$ ($26.0$) and $4530$ ($66.0$) galaxies (per cent) are classified as voids, sheets and  filaments/knots. The fractions of void and sheet galaxies hardly change from the quaternary classification. In the binary classification, $927$ ($13.5$) and $5939$ ($86.5$) galaxies (per cent) are classified as voids, sheets/filaments/knots. Although the fraction of voids is somewhat higher than those in the cases of the quaternary and ternary classification, the binary case is expected to be more accurate and credible for the voids because of the higher $F_{\rm 1}$-score (Fig. \ref{CM_galbaseObs100Mpc_all}).

\section{Discussion}
\label{discussion}

\subsection{Classification with local density of galaxies}
\label{locden}
\begin{figure}
  \includegraphics[bb=0 0 1853 2207, width=\hsize]{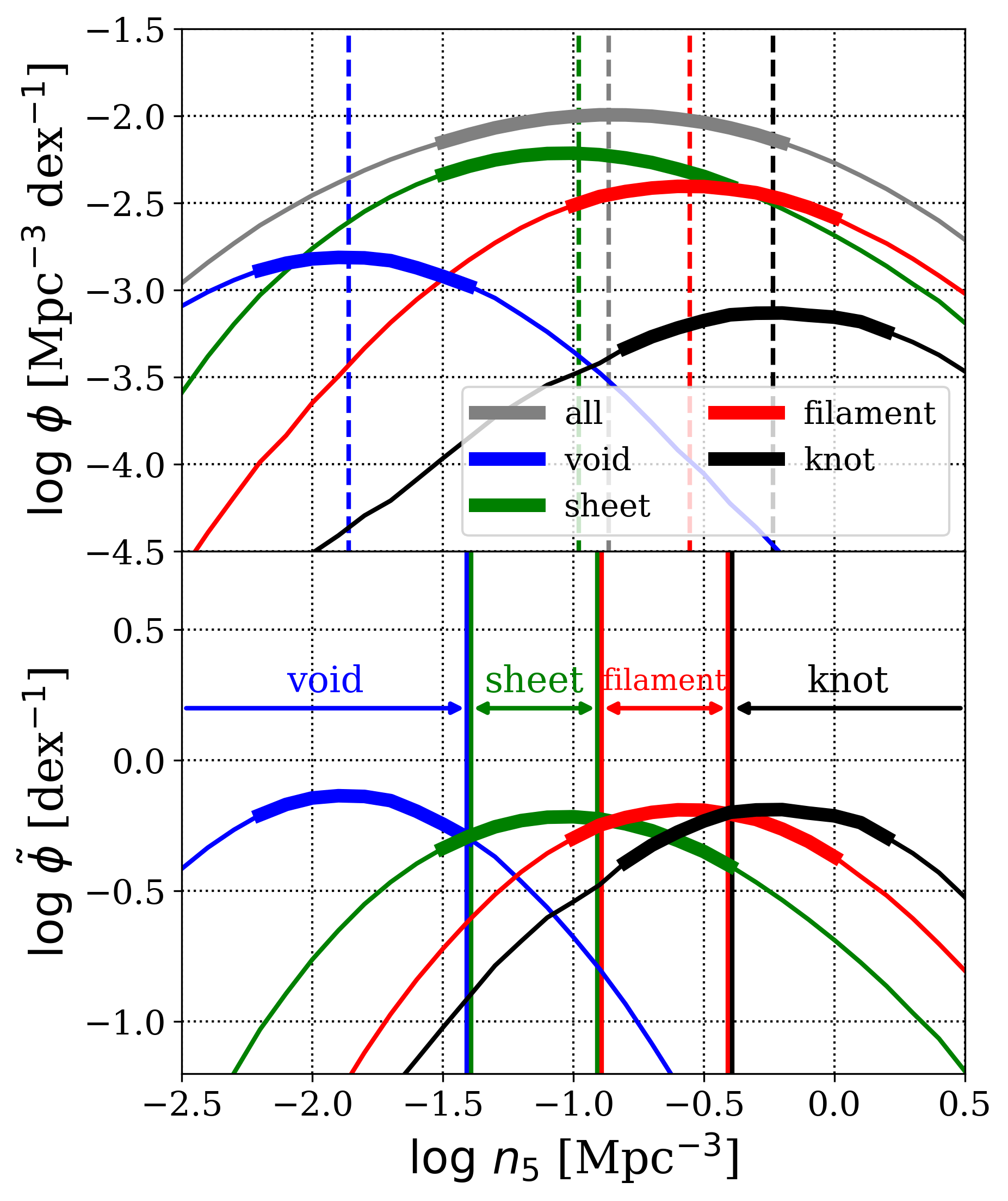}
\caption{\textit{Top:} Global number density $\phi$ of galaxies in the whole simulation as functions of local number density $n_5$ of galaxies within a radius of the fifth nearest neighbour. Here the galaxies brighter than $M_r=-17.25~{\rm mag}$ are included in the analysis, and their line-of-sight positions are affected by $v_{\rm los}$ according to equation (\ref{losv}). The vertical dashed line indicates the median of $n_5$ for each (true) class label and all the galaxies, and the thick part of the solid line highlights the range of $\pm1\sigma$ from the median. \textit{Bottom:} Same as the top panel but normalised to make the integral unity: $\int\tilde{\phi} \mathrm{d}n_5=1$. The vertical solid lines indicate $n_5$ at which the probability density functions $\tilde{\phi}$ become equal between the contiguous classes. We consider these values of $n_5$ as the boundaries between the classes and give the galaxies their predicted labels as function of $n_5$. The allows illustrate the ranges of $n_5$ for the predicted classes.}
\label{k5Density}
\end{figure}

Here, we compare the accuracy of our 3D-CNN model with that of a simple method not using the deep learning. In the four cosmic-structure classes, local densities of galaxies are expected to typically increase from void, sheet, filament to knot regions. A local density around a galaxy may therefore characterise the cosmic structure to which the galaxy belongs. As is done in Section \ref{withobsrest}, we exclude galaxies fainter than $M_r=-17.25~{\rm mag}$ from the TNG100-1 simulation and shift positions of the galaxies according to their $v_{\rm los}$ by equation (\ref{losv}). We then compute a distance from a galaxy to its fifth nearest neighbour, $r_5$. We calculate a local number density of galaxies as $n_5=5/(4\upi r_5^3/3)$ for every single galaxy with $M_r<-17.25~{\rm mag}$ in the simulation, compute a global number density $\phi$ of galaxies with a given $n_5$ in the entire simulation. The top panel of Fig. \ref{k5Density} shows the distribution of $\phi$ as a function of $n_5$ for galaxies in each (true) class label. As we expect above, the median values of $n_5$ (the vertical dashed lines) increase from voids to knots in the order. It is worthy to mention, however, that their ranges between $\pm1\sigma$ of $n_5$ (the thick parts of the solid lines) significantly overlap with each other, especially those of sheets, filaments and knots.\footnote{Similar arguments have been presented in \citet{hmy:12} for DM density.} We find that the distance measurement errors by $v_{\rm los}$ can significantly lower $n_5$ of galaxies in dense environments.

In the bottom panel of Fig. \ref{k5Density}, we normalise the functions of $\phi$ to make their integrals equal to unity. The vertical solid lines indicate the values of $n_5$ at which the normalised distribution functions $\tilde{\phi}$ become equal between contiguous classes. We consider these values of $n_5$ to be the thresholds between the pairs of contiguous classes, which give the galaxies their predicted labels as illustrated in the bottom panel of Fig. \ref{k5Density}. Thus, in this simple analysis, each galaxy has its true label determined by the DM analysis (equation \ref{VelGraTendor}) and the predicted label given by the local density $n_5$. A difference from the 3D-CNN models is that this prediction based on $n_5$ does not take into account the shapes of three-dimensional distribution of galaxies. 

\begin{figure}
  \includegraphics[bb=0 0 769 493, width=\hsize]{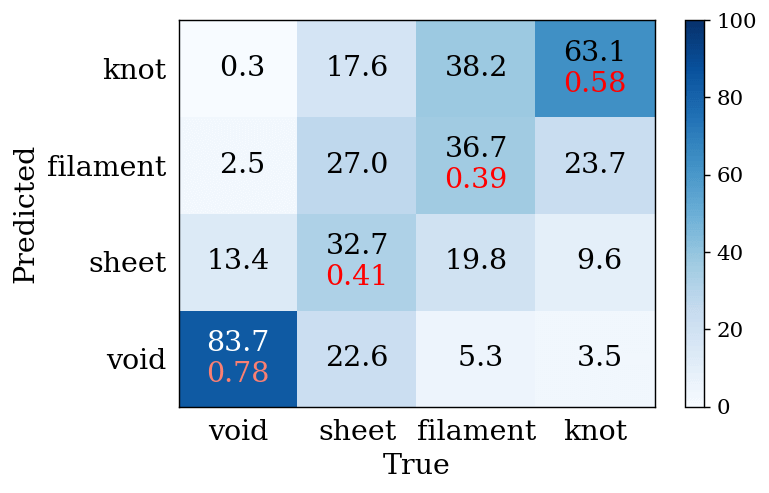}
\caption{Same as the top panel of Fig. \ref{CM_galbaseObs100Mpc_all} but for the predicted classes given by the local density $n_5$ described in Section \ref{locden} and Fig. \ref{k5Density}. The macro-average of the $F_{\rm 1}$-scores is $0.54$, and the number-averaged $F_{\rm 1}$-score is $0.44$.}
\label{CM_localdensity}
\end{figure}
Fig. \ref{CM_localdensity} shows the normalised confusion matrix of the above classification. In comparison with the 3D-CNN model (the top panel of Fig. \ref{CM_galbaseObs100Mpc_all}), the $F_{\rm 1}$-scores decrease in all classes, and their macro-average is $0.54$. Especially, the misclassification of filaments as knots ($38.2$ per cent) is more than the true positive classification of filaments ($36.7$ per cent). However, the decrease of the $F_{\rm 1}$-scores is relatively small in the voids and knots. Although we define $n_5$ at the fifth nearest neighbour, the result of Fig. \ref{CM_localdensity} does not significantly depend on the number of the nearest galaxies to define the local density. If predicted labels are given completely at random, quaternary classification results in $F_{\rm 1}=0.25$ for all classes. So, in this sense, we consider that the classification based on $n_5$ is not too inaccurate. Although the local density of galaxies is an essential quantity to characterise the cosmic structures, the 3D-CNN model can improve the accuracy by taking into account the three-dimensional extension of galaxy distribution, especially for sheets and filaments.\footnote{If the observational restrictions are ignored, we find that the simple classification method based on the local density $n_5$ does not classify the simulated galaxies accurately. This is because a large number of faint galaxies are taken into account in this analysis due to the absence of the limiting magnitude, and the number density of galaxies increases generally. Since most of the faint galaxies are satellites, the distances to the fifth nearest neighbours $r_5$ can become so short that the fifth neighbours can be included in the systems hosting the galaxies (i.e. $r_5\sim10$--$100~{\rm kpc}$). Since the local density $n_5$ is strongly affected by their host systems in this case, $n_{\rm 5}$ cannot characterise their environments of the cosmic structures on the scale of $\sim10~{\rm Mpc}$. Even if we use the fiftieth nearest neighbours $n_{50}$, the simple classification method is still significantly inaccurate.}

a large number of faint galaxies are taken into account in this analysis, and the number density of galaxies increases generally. Since most of the faint galaxies would be satellites, the distances to the fifth nearest neighbours $r_5$ can become significantly shorter.

\subsection{To improve the performance}
\label{improve}

We have tested various types of CNN architectures. Two-dimensional CNN using projected images of the cubic data are significantly inaccurate. More sophisticated 3D-CNN on neither Residual Networks \citep[ResNet, ][]{ResNet} nor U-Net based classification models improve the accuracy significantly. We finally arrived at the simple 3D-CNN classifier described in Section \ref{CNN}. 

In the cosmological simulation, there are other quantities available in observations besides the number of galaxies, such as stellar mass and colour. As the second and third channels of our cubic data, we add the total stellar mass and $g-i$ colours of galaxies in each voxel. However, we find that these additional channels do not change the performance of our models. Although we have also tested with the higher-resolution simulation of TNG50-1, the results are similar to those with TNG100-1 (see Appendix \ref{TNG50}). 

The method of this study assumes the cosmological simulation to be accurate and statistically compatible with the observed galaxies. The details of baryon components can depend on the sub-grid models in the simulation. However, since our fiducial models only use the spatial distribution of galaxies above the limiting magnitude, our results are expected to be less dependent on such physical models for baryons. Our results can also depend on the accuracy of the cosmic-structure classification to obtain the true labels. In Section \ref{sim_csc}, we have used the method of \citet{hmy:12} with the threshold of $\lambda_{\rm th}=0.44$. If this method and/or the threshold is inaccurate, our models cannot correctly predict the labels by the mislabelling. In addition, the eigenvalues of the velocity-gradient tensors (equation \ref{VelGraTendor}) can vary continuously between the classes. If it is the case, the class labels given to a certain fraction of galaxies would be ambiguous. However, there are no methods to evaluate quantitatively the accuracy of the true labels. In this study, we consider that the DM analysis of \citet{hmy:12} presents the `definitions' of the cosmic structures. The detailed discussion of the cosmic-structure classification is beyond the scope of this study although we show further analysis for the true labels in Appendix \ref{mislabel}.

\begin{figure}
  \includegraphics[bb=0 0 769 493, width=\hsize]{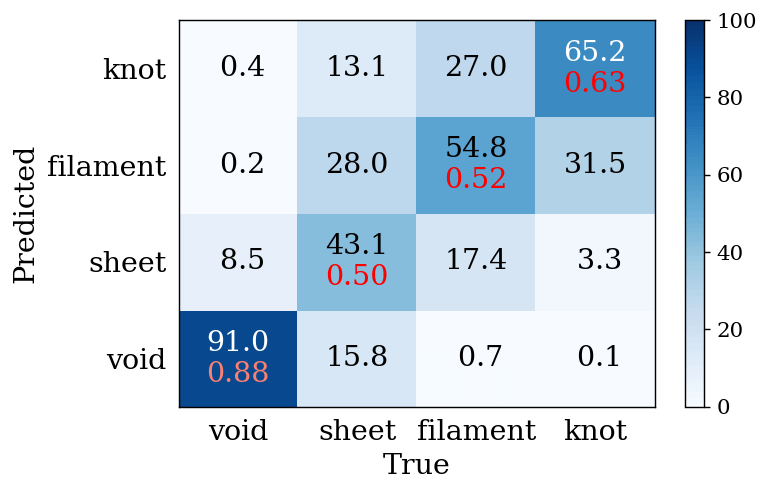}
\caption{Same as the top panel of Fig. \ref{CM_galbaseObs100Mpc_all} but using all haloes having stellar particles by ignoring the magnitude limit. Their line-of-sight positions are affected by $v_{\rm los}$. The macro-average of the $F_{\rm 1}$-score is $0.64$ for the test data, and the number average is $0.54$.}
\label{CM_NoLimMag}
\end{figure}
Observations would be improved in the future. In Section \ref{withobsrest}, we have considered the two observational restrictions: limiting magnitude and distance measurement error by a proper motion. The former can be mitigated if deeper spectroscopic surveys become available although the latter is unavoidable as long as distances are measured from recession velocities of galaxies. Fig. \ref{CM_NoLimMag} shows the confusion matrix of the galaxy-based classification, where the limiting magnitude is ignored by taking into account all the subhaloes that contain stellar particles whereas the distance errors are included. In comparison with the result including both restrictions (the top panel of Fig. \ref{CM_galbaseObs100Mpc_all}), the performance of the model is a little improved in Fig. \ref{CM_NoLimMag}, and the macro-averaged $F_{\rm 1}$-score is $0.64$. This $F_{\rm 1}$-score is similar to that in the case including neither restriction (the top panel of  Fig. \ref{CM_galbase_all}). Thus, although excluding faint galaxies by the limiting magnitude lowers the accuracy of the models, it can be improved if the limiting magnitude is mitigated in future observations. In addition, we evaluate the influence by the distance errors due to $v_{\rm los}$ by comparing Fig. \ref{CM_NoLimMag} with the top panels of Fig. \ref{CM_galbase_all}. Although the model for Fig. \ref{CM_NoLimMag} includes the effect of $v_{\rm los}$, the $F_{\rm 1}$-scores are similar to those in the top panels of Fig. \ref{CM_galbase_all}, and the differences are within the fluctuation shown in Table \ref{repeat} (Appendix \ref{inbalance}). This may imply that the 3D-CNN model learns the pattern of the redshift distortions by $v_{\rm los}$ in cluster regions, and the model can adjust the prediction to the distance errors. Thus, the distance errors do not make the models inaccurate.

\section{Summary and conclusions}
\label{summary}
This study explore the ability of deep learning to classify the cosmic structures into four categories: knot, filament, sheet and void. Utilising the cosmological simulation, these labels are obtained from local velocity-gradient tensors of DM, and our models based on 3D-CNN predict the labels from spatial positions of galaxies within a cubic region of $10^3~{\rm Mpc^3}$ centred on a classification point. We train the models with data of the cosmological simulation, which can include observational restrictions such as limiting magnitude and distance measurement error by a proper motion of a galaxy. Since the input data are distribution of observable galaxies, our 3D-CNN model can be directly applied to observations of galaxy surveys such as SDSS. Thus, our models can tag the observed galaxies with the class labels obtained from the DM-based analysis in the simulation. In this sense, the method proposed in this paper is qualitatively different from the previous classification for observed galaxies. Our approach connects cosmological simulations to observations with the aid of deep learning.

If the classification points are randomly selected from uniform spatial grid points, our model can achieve the macro-averaged $F_{\rm 1}$-score of $0.74$ in the quaternary classification. It is generally difficult to distinguish sheets and filaments, whereas voids and knots can be classified with relatively high accuracy. In the case of the binary classification into sheet and filament/knot, the macro-averaged $F_{\rm 1}$-score is $0.84$. This accuracy is comparable to the result of \citet{a:19} in which learning data for his similar 3D-CNN model are created from DM density fields, whereas our model uses galaxy distribution. This means that galaxy distribution can be a substitution for DM density fields, and wide-field observations of galaxy surveys can be used to classify the cosmic structures by the 3D-CNN models.

To classify galaxies, we need to select the classification points from the positions of galaxies in the simulation. In this case, the macro-averaged $F_{\rm 1}$-score is $0.64$ in the quaternary classification. If we include distance measurement errors by proper motions and impose the limiting magnitude of the SDSS spectroscopy, our model results in the averaged $F_{\rm 1}$-score of $0.60$ in the quaternary classification. For the specific purpose to identify void galaxies, the $F_{\rm 1}$-score reaches the high value of $0.88$ in the binary classification distinguishing between voids and the others. These deep-learning models can be applied to SDSS data, and the results are shown in Fig. \ref{SDSS_classification}.

\section*{Acknowledgements}
We are grateful to the reviewer, Miguel Aragon-Calvo, for his useful comments and suggestion. We thank Yoshihiro Takeda, Masami Ouchi and Hidenobu Yajima for their fascinating discussion and useful suggestion. Numerical computations were in part carried out on GPU/GRAPE system and analysis servers at Center for Computational Astrophysics, National Astronomical Observatory of Japan.
This work is supported by MEXT/JSPS KAKENHI Grant Numbers (JP19H01931, JP20H05861, JP21H04496).

\section*{Data Availability}
The data underlying this article will be shared on reasonable request to the corresponding author.



\bibliographystyle{mnras}




\appendix
\section{The limiting magnitude in SDSS}
\label{SDSSlimit}
\begin{figure}
  \includegraphics[bb=0 0 1760 703, width=\hsize]{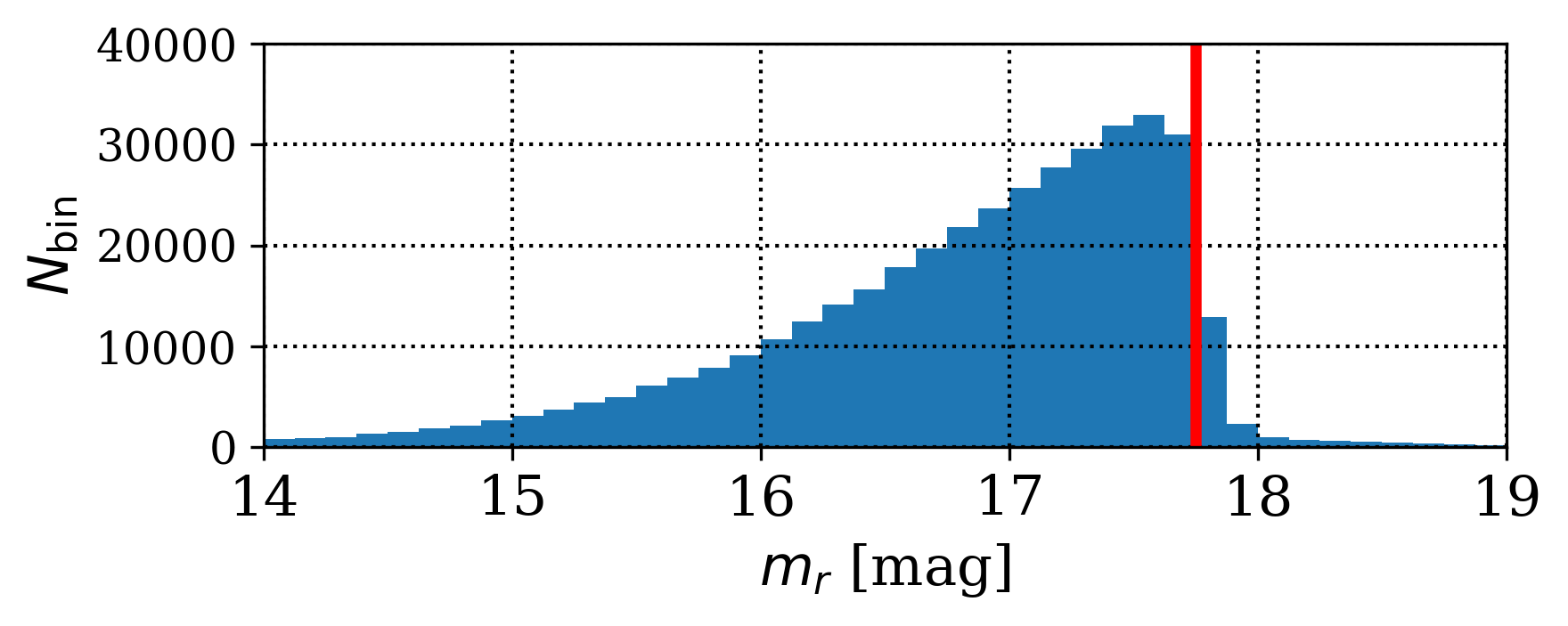}
\caption{The histogram of $r$-band apparent magnitudes of the SDSS galaxies with distances $<500~{\rm Mpc}$. The vertical red line indicates the limiting magnitude $m_r=17.75$ below which 95 per cent of the galaxies are included.}
\label{SDSShist}
\end{figure}
For the models used in Sections \ref{withobsrest} and \ref{appSDSS}, we need to make the galaxy samples consistent between the simulation and SDSS. For this purpose, we determine the limiting magnitude of the SDSS spectroscopy. From the catalogue of SDSS DR12 for the north side, we extract galaxies and quasars with redshift measurements by spectroscopy and plot the distribution of their $r$-band apparent magnitude $m_r$ in Fig. \ref{SDSShist}. The distribution of $m_r$ sharply declines at $m_r=17.75$ indicated with the vertical red line, and 95 per cent of the galaxies within $d_{\rm obs}=500~{\rm Mpc}$ have $m_r<17.75$. Thus, $m_r=17.75$ is considered to be the limiting magnitude of the spectroscopy for their redshift measurements. We find that the limit can be determined the most accurately in the distribution of $r$-band magnitudes. In Section \ref{withobsrest}, we adopt this limit to the simulation and exclude galaxies fainter than the limit from our sample. Imposing the limit significantly decreases the number of simulated galaxies (see Fig \ref{HOF} and Section \ref{withobsrest}).

\section{Results for TNG50-1}
\label{TNG50}
The run of TNG50-1 has the highest mass-resolution in the series of the IllustrisTNG simulations. The simulation box has a side length of $51.7~{\rm Mpc}$, and DM and stellar particles in TNG50-1 have $8.5\times10^4$ and $4.5\times10^5~{\rm M_\odot}$ which are $16.5$ times smaller than those in TNG100-1 used in our fiducial cases. Therefore, TNG50-1 samples a larger (smaller) number of faint (bright) galaxies than TNG100-1. In the snapshot data at redshift $z=0$, we apply the same classification analysis of \citet{hmy:12} to DM with a resolution of $\Delta r=202~{\rm kpc}$ ($256^3$ voxels). Galaxies having stellar particles occupy most of haloes above $\sim10^{8.5}~{\rm M_\odot}$, and $14211$, $75924$, $46110$ and $9687$ galaxies are labelled as voids, sheets, filaments and knots, respectively. However, when we impose the limiting magnitude of $M_r=-17.25$, the numbers of galaxies decrease to $158$, $2318$, $1374$ and $324$. The small sample size is thought to be insufficient to create 10000 cubic data for each class, especially for voids and knots. Accordingly, we consider TNG50-1 to be unsuitable for the 3D-CNN models including the observational restrictions (Section \ref{withobsrest}).

\begin{figure}
  \includegraphics[bb=0 0 769 493, width=\linewidth]{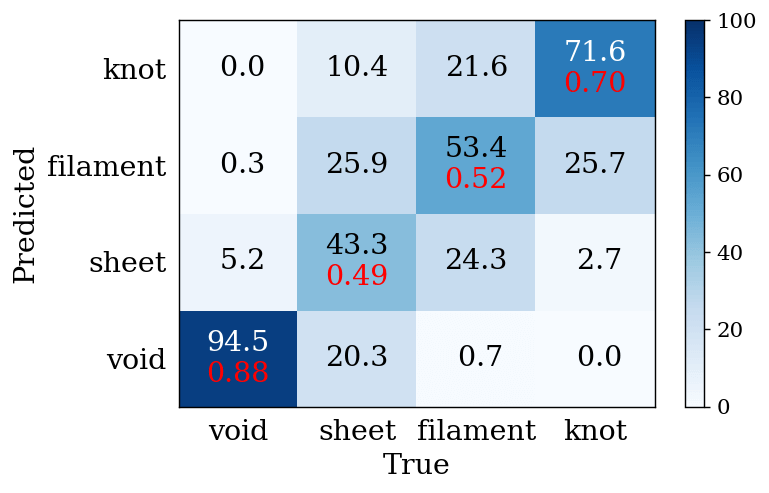}
  \includegraphics[bb=0 0 769 524, width=\linewidth]{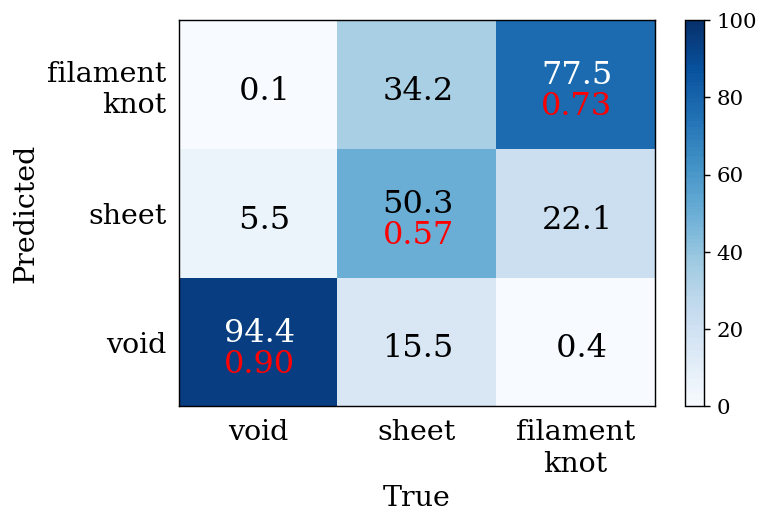}
  \includegraphics[bb=0 0 769 555, width=\linewidth]{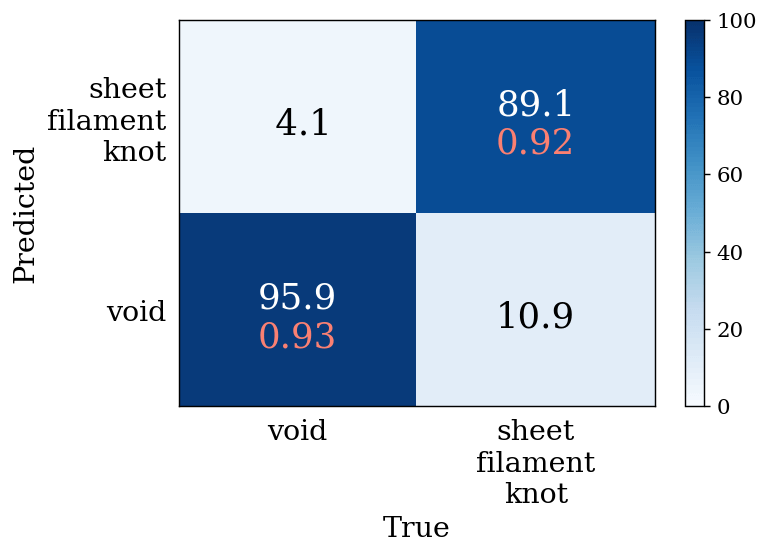}
\caption{Same as \ref{CM_galbase_all} but for TNG50-1: classifying the galaxy-based data cubes created without the observational restrictions. The macro-averages of $F_{\rm 1}$-scores among the classes are $0.66$, $0.73$ and $0.93$ in the top, middle and bottom panels, whereas the `number-averaged' $F_{\rm 1}$-scores are $0.55$, $0.78$ and $0.92$.}
\label{CM_TNG50}
\end{figure}
As is done in Section \ref{noobsrest}, we construct the models for the galaxy-based classification without considering the observational restriction. The cubic data for learning are created in the same way as described in Sections \ref{CreateData} and \ref{DataAug} but with a smaller side length of $5~{\rm Mpc}$ for the volume of the cubic data due to the smaller simulation box and higher resolution of TNG50-1.\footnote{We confirm that our result for TNG50-1 hardly changes even if the side length of the cubic data is kept to be $10~{\rm Mpc}$.} Fig. \ref{CM_TNG50} shows the normalised confusion matrixes and $F_{\rm 1}$-scores for the models with TNG50-1. In the  macro-average values, the $F_{\rm 1}$-scores are higher than in the fiducial case with TNG100-1 by $0.02$ and $0.01$ in the quaternary and binary classification but lower by $0.01$ in the ternary classification. Thus, using the higher-resolution run TNG50-1 does not significantly improve the performance of our 3D-CNN models.

\section{The influence by the analysis for the true labels}
\label{mislabel}
In the bottom panel of Fig. \ref{DMmap}, some weak structures of sheets appear to be discontinuous and/or hollow in their `core' regions that are categorised as voids. These features may be indicative of the inaccuracy of the DM analysis based on \citet{hmy:12}; such discontinuous and hollow cores of sheets are also seen in \citet[][their fig. 1]{hmy:12}. The presence of the discontinuous cores can be explained physically in the analysis of \citet{hmy:12}. DM density is not uniform in a sheet and decreases towards its centre. Since the central core region is expected to have the lowest density in the sheet, the density contrast between the core and neighbouring void regions is weak. The local environment of such a core is thought to be effectively the same as those of the surrounding void regions. Since the potential well is shallow in the core, the velocity gradient (the largest eigenvalue of $\Sigma_{ij}$ in equation \ref{VelGraTendor}) becomes small in/around the core. The fiducial threshold of $\lambda_{\rm th}=0.44$ would be so strict that the analysis cannot capture such a low-density core in the sheet. We expect, however, that there are few galaxies in such low-density environments, and labelling the sheet cores as voids would not significantly affect our results in the galaxy-based classification.

A possible remedy for the discontinuity of sheets is to adopt a lower threshold in the cores. We here present our new analysis for the true labels, where we first compute the labels with the fiducial threshold $\lambda_{\rm th}=0.44$ and iterate the same computations but with a secondary threshold $\lambda_{\rm th}'$ for the void regions detected first with the fiducial threshold. If the results change in the re-computations with $\lambda_{\rm th}'$, we replace the labels. Although we infer that the hollow parts of sheets would be attributed to the over-resolution in the spines of the sheets, we find that the hollowness is also mitigated by the above re-labelling.

\begin{figure}
  \includegraphics[bb=0 0 1476 1282, width=\linewidth]{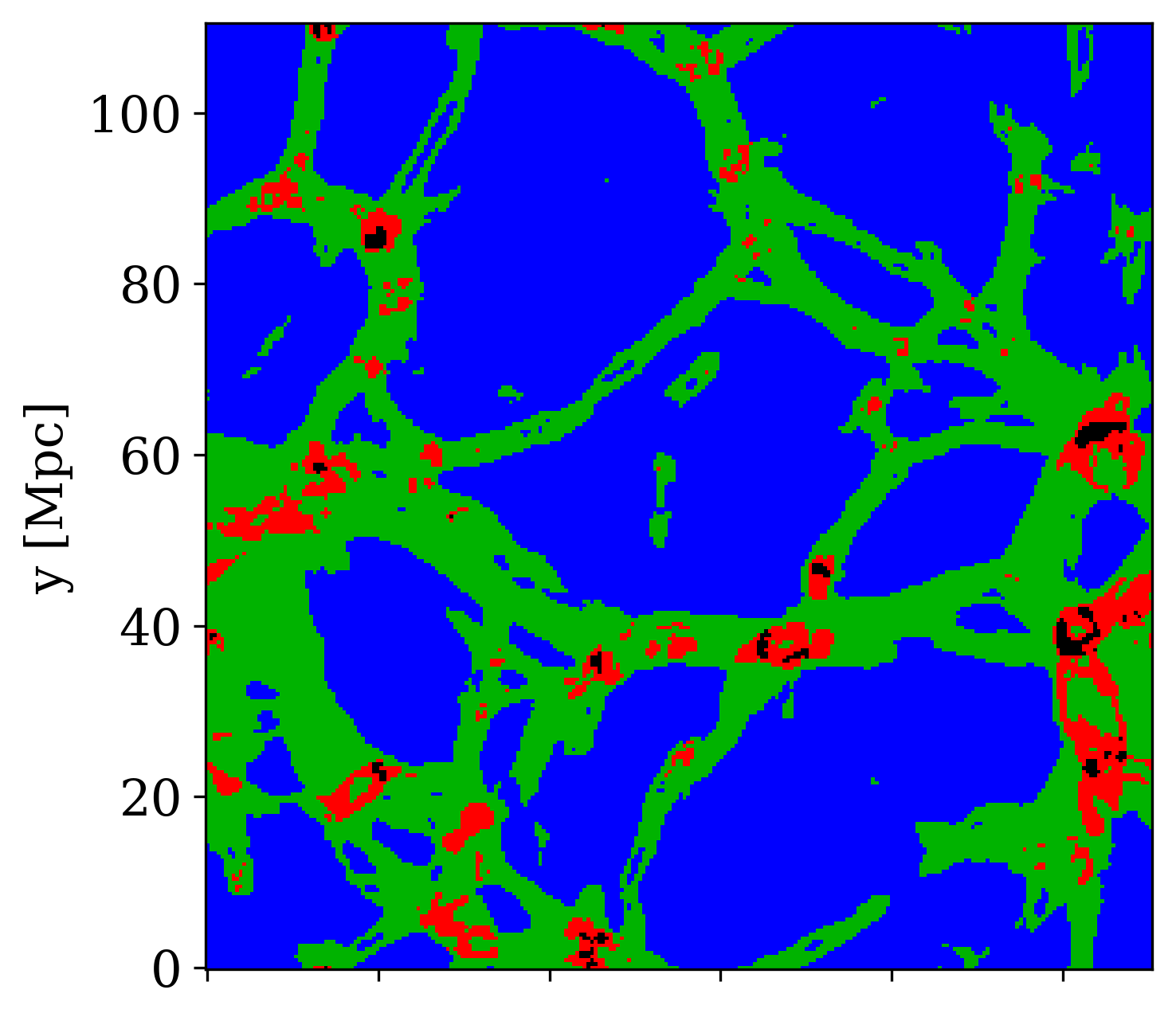}
  \includegraphics[bb=0 0 1476 1398, width=\linewidth]{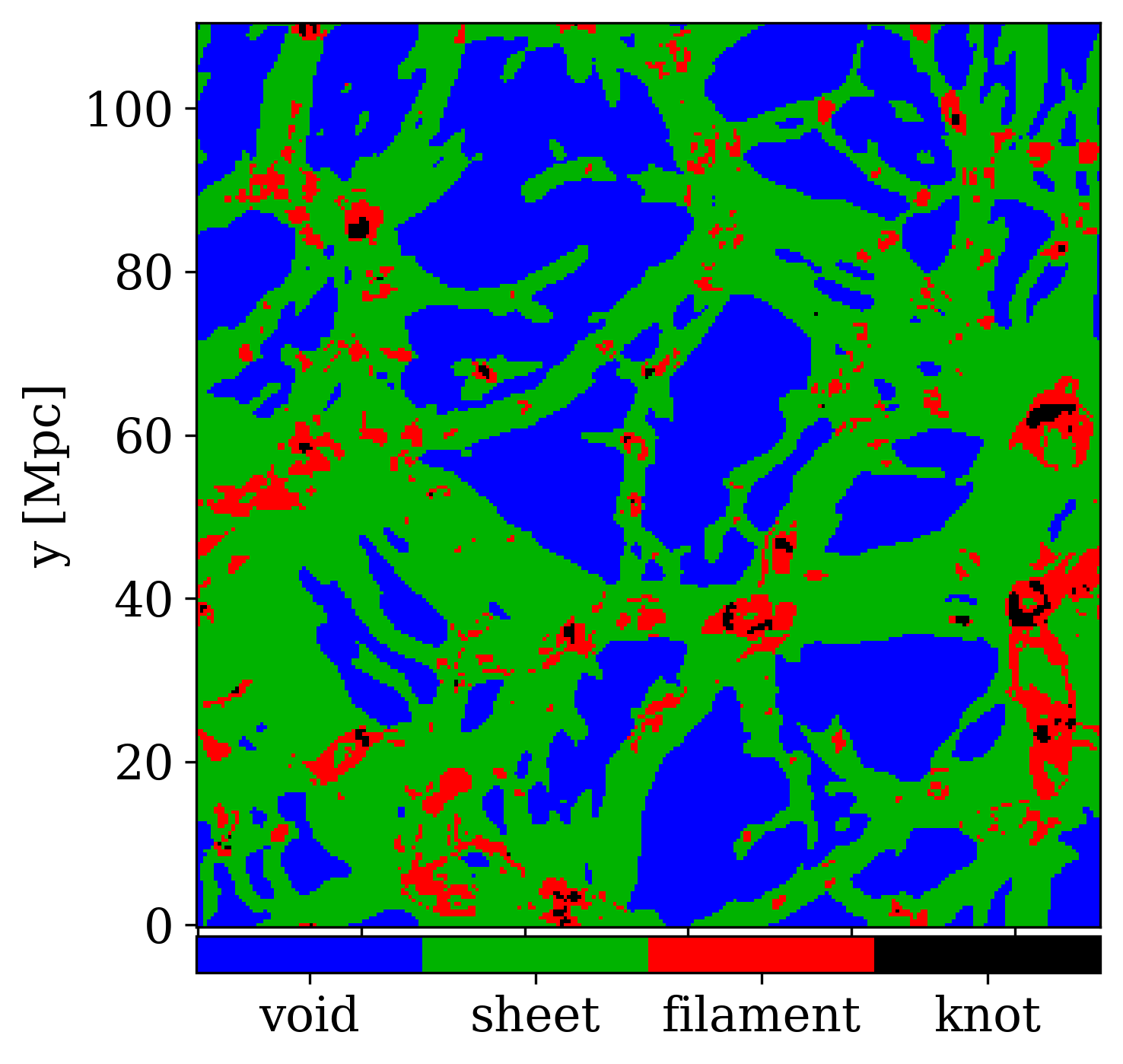}
\caption{Same as the bottom panel of Fig. \ref{DMmap}, but the re-labelling process is applied to the regions first identified as void with the fiducial threshold $\lambda_{\rm th}=0.44$. In the top and bottom panels, the re-labelling is performed with the lower secondary thresholds $\lambda_{\rm th}'=0.30$ and $0.05$.}
\label{RelabellingMaps}
\end{figure}
Fig. \ref{RelabellingMaps} shows the cosmic-structure classification with the re-labelling for voids with the secondary thresholds $\lambda_{\rm th}'=0.30$ and $0.05$ for the same slice of Fig. \ref{DMmap} in Section \ref{sim_csc}. As $\lambda_{\rm th}'$ decreases, the sheets become thicker, and the discontinuous and/or hollow features seem to be improved to some extent. Note, however, that even weaker structures in low-density environments are newly identified as sheets, and some of them are still discontinuous and/or hollow. It should also be noted that the re-labelling process with $\lambda_{\rm th}'$ turns void regions into sheets/filaments not only in the sheet cores but also in other regions close to sheets. From appearance of the bottom panel of Fig. \ref{RelabellingMaps}, the secondary threshold of $\lambda_{\rm th}'=0.05$ may be too low since the sheet regions occupy a larger volume than the voids. In the grid-based classification with the secondary thresholds of $\lambda_{\rm th}'=0.30$ ($0.05$), $59$ ($30$), $36$ ($61$), $4.9$ ($8.1$) and $0.34$ ($0.40$) per cent of the spatial voxels are categorised as void, sheet, filament and knot, respectively. In the galaxy-based classification with the secondary thresholds of $\lambda_{\rm th}'=0.30$ ($0.05$), if we do not take into account the observational restrictions, 25013 (1293), 173299 (168169), 105514 (130856) and 21907 (25415) galaxies are labelled as voids, sheets, filaments and knots, respectively. The number of void galaxies monotonically decreases as $\lambda_{\rm th}'$ decreases, whereas the number of sheet galaxies hardly changes with $\lambda_{\rm th}'$.

\begin{figure}
  \includegraphics[bb=0 0 769 493, width=\linewidth]{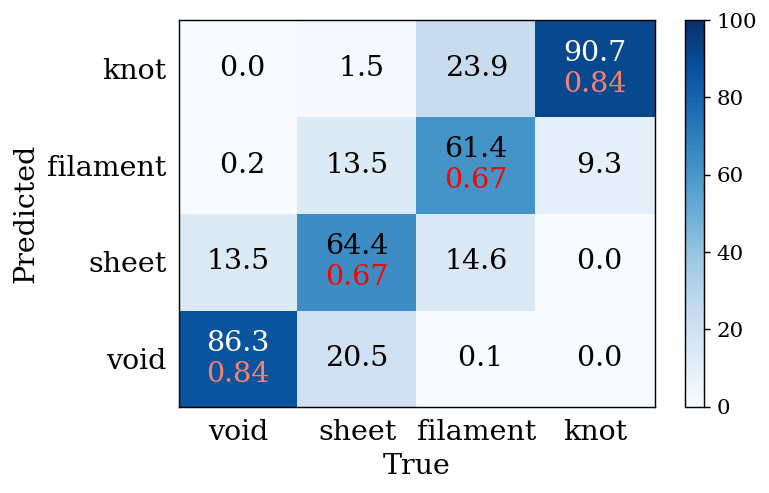}
  \includegraphics[bb=0 0 769 493, width=\linewidth]{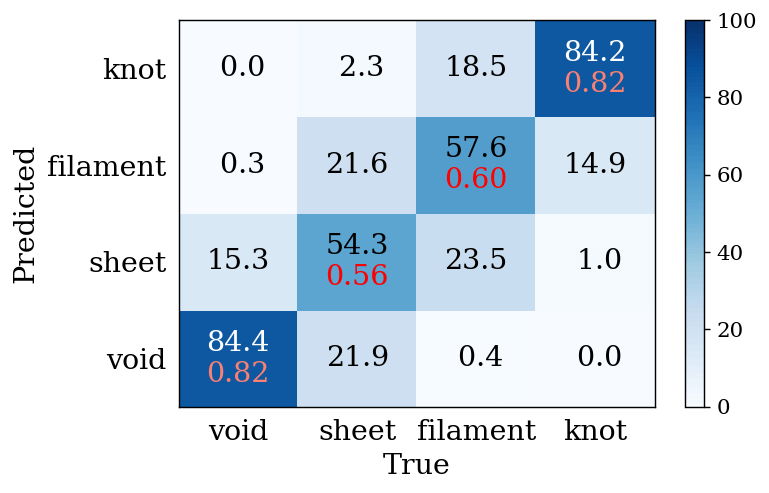}
\caption{Confusion matrixes and $F_{\rm 1}$-scores in the grid-based classification where the true labels are obtained the DM analysis with the re-labelling with $\lambda_{\rm th}'=0.30$ (top) and $0.05$ (bottom). The macro-averaged $F_{\rm 1}$-scores are $0.75$ and $0.70$ for $\lambda_{\rm th}'=0.30$ and $0.05$.}
\label{RelabellingCM_Grid}
\end{figure}
\begin{figure}
  \includegraphics[bb=0 0 769 493, width=\linewidth]{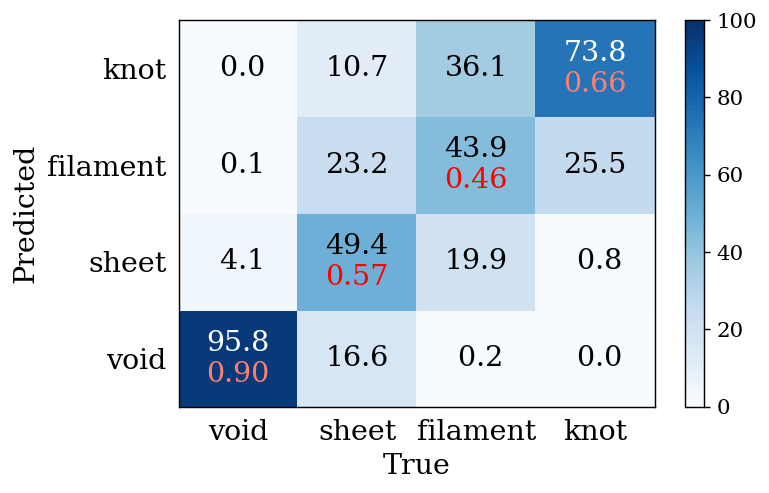}
  \includegraphics[bb=0 0 769 493, width=\linewidth]{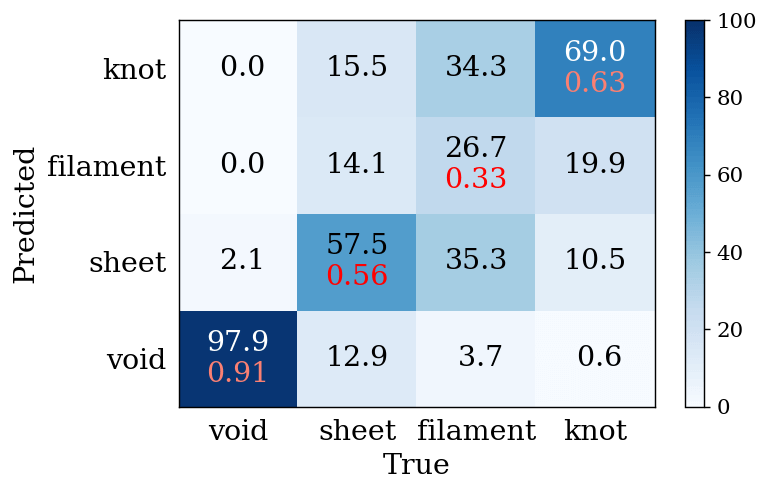}
\caption{Same as Fig. \ref{RelabellingCM_Grid} but for the case of the galaxy-based classification without the observational restrictions. The macro-averaged $F_{\rm 1}$-scores are $0.65$ and $0.61$ for $\lambda_{\rm th}'=0.30$ (top) and $0.05$ (bottom).}
\label{RelabellingCM_Gal}
\end{figure}
Figs. \ref{RelabellingCM_Grid} and \ref{RelabellingCM_Gal} show the results of our 3D-CNN models using the true labels obtained by the above DM analysis with the re-labelling. In the grid-based classification (Fig. \ref{RelabellingCM_Grid}), the macro-averages of $F_{\rm 1}$-scores are $0.75$ and $0.70$ for $\lambda_{\rm th}'=0.30$ (top) and 0.05 (bottom). Since the macro-averaged $F_{\rm 1}$-score is $0.74$ in the fiducial case (Fig. \ref{CM_grid_four} in Section \ref{gridbase}), the re-labelling with $\lambda_{\rm th}'$ does not appear to affect the performance of our model. The low value of $\lambda_{\rm th}'=0.05$ appears to make the prediction inaccurate slightly. In the galaxy-based classification without the observational restrictions (the top panel of Fig. \ref{RelabellingCM_Gal}), the macro-averages of $F_{\rm 1}$-scores are $0.65$ and $0.61$ for the re-labelling with $\lambda_{\rm th}'=0.30$ (top) and 0.05 (bottom). We obtain the same result: the re-labelling little affects the performance of our 3D-CNN model. The slightly lower macro-averaged $F_{\rm 1}$-scores in the cases of $\lambda_{\rm th}'=0.05$ could be attributed to the significance of mis-labelling due to the low secondary threshold. From the above results, we consider that the fiducial classification based on \citet{hmy:12} is reasonable enough, and our model can be readily applied to true labels classified by other methods since the performance of the models hardly depends on the details of the DM analysis to obtain the true labels.

\section{The influence by the class inbalance}
\label{inbalance}
The class inbalance is large in our original samples taken from the entire simulation (see Section \ref{sim_csc}). In the grid-based classification, the majority of the voxels in the whole simulation are categorised as voids, whereas the knot regions are quite small. In the galaxy-based classification, the majority of the galaxies are (true) labelled as sheets and filaments. We here argue the influence by the class inbalance on our results.

\begin{table}
  \begin{center}
    \leavevmode
\begin{tabular}{|l | c c c c c |c} 
\hline
\hline
           & void & sheet & filament & knot &macro-average\\
\hline 
minimum    & 0.8260 & 0.6216 & 0.6127 & 0.8229 & 0.7311 \\
$-1\sigma$ & 0.8289 & 0.6326 & 0.6308 & 0.8258 & 0.7365 \\
median     & 0.8399 & 0.6497 & 0.6360 & 0.8329 & 0.7405 \\
$+1\sigma$ & 0.8492 & 0.6768 & 0.6554 & 0.8379 & 0.7468 \\
maximum    & 0.8495 & 0.6786 & 0.6671 & 0.8411 & 0.7522 \\
\hline 
  \end{tabular} 
\begin{tabular}{|l | c c c c c |c} 
\hline
\hline
           & void & sheet & filament & knot &macro-average\\
\hline 
minimum    & 0.8779 & 0.5033 & 0.3826 & 0.6494 & 0.6202 \\
$-1\sigma$ & 0.8829 & 0.5046 & 0.4094 & 0.6618 & 0.6295 \\
median     & 0.8880 & 0.5411 & 0.4390 & 0.6796 & 0.6347 \\
$+1\sigma$ & 0.8932 & 0.5705 & 0.4573 & 0.6895 & 0.6427 \\
maximum    & 0.8970 & 0.5779 & 0.4763 & 0.6925 & 0.6443 \\
\hline 
\hline
  \end{tabular} 
\caption{Statistics of $F_{\rm 1}$-scores in our 3D-CNN models by repeating the same computations 10 times with different random seeds. \textit{Top:} the grid-based classification. \textit{Bottom:} the galaxy-based classification without the observational restrictions.}
  \label{repeat}
  \end{center}
\end{table}
Because we randomly create the cubic data equally in number between the class labels (10000 for each, and 6400 of them are used for the training) in all cases, the class inbalance does not directly affect the training of our 3D-CNN models. Sampling 10000 cubic data may, however, be insufficient for the large classes such as void in the grid-based classification and filament and sheet in the galaxy-based classification. If it is the case, our 3D-CNN models could not be robust. To evaluate the influence, we here examine the results of the grid-based quaternary classification (Fig. \ref{CM_grid_four}) and the galaxy-based quaternary classification without the observational restrictions (the top panel of Fig. \ref{CM_galbase_all}). We repeat the same computations 10 times while changing random seeds in creating the cubic data. Table \ref{repeat} shows the statistics of $F_{\rm 1}$-scores among the 10 repeated computations. In both of the grid- and galaxy-based classification, the fluctuations of $F_{\rm 1}$-scores are larger in the classes of sheets and filaments than those in voids and knots. Especially, the $F_{\rm 1}$-scores of filament galaxies can vary by $\sim0.1$ from the minimum to maximum in the galaxy-based classification. However, the macro-averages of $F_{\rm 1}$-scores only fluctuate by $\sim0.02$.


\bsp	
\label{lastpage}
\end{document}